\documentclass[prc,twocolumn,floatfix,groupedaddress,nofootinbib,preprintnumbers,amsmath,amssymb,amsfonts,superscriptaddress,widetable,href]
{revtex4}
\usepackage{amsmath,amssymb,amsfonts}
\usepackage{ulem}
\usepackage{graphicx}
\usepackage{color}
\usepackage{longtable}

\newcommand{\be}{\begin{equation}}
\newcommand{\ee}{\end{equation}}
\newcommand{\bea}{\begin{eqnarray}}
\newcommand{\eea}{\end{eqnarray}}

\begin{document}

\title{\bf Light projectile elastic scattering by nuclei described by the Gogny interaction}

%
%
%
\author{J. L\'{o}pez Mora\~na}
\affiliation{Departament de F\'isica Qu\`antica i Astrof\'isica (FQA),	Universitat de Barcelona (UB), Mart\'i i Franqu\`es 1, E-08028 Barcelona, 
Spain}
\affiliation{Institut de Ci\`encies del Cosmos (ICCUB),	Universitat de Barcelona (UB), Mart\'i i Franqu\`es 1, E-08028 Barcelona, Spain}
\author{X. Vi\~nas}
\affiliation{Departament de F\'isica Qu\`antica i Astrof\'isica (FQA), Universitat de Barcelona (UB), Mart\'i i Franqu\`es 1, E-08028 Barcelona, 
Spain}
\affiliation{Institut de Ci\`encies del Cosmos (ICCUB), Universitat de Barcelona (UB), Mart\'i i Franqu\`es 1, E-08028 Barcelona, Spain}
\affiliation{Institut Menorqu\'i d'Estudis, Cam\'i des Castell 28, 07702 Ma\'o, Spain}
%

\date{\today}

\begin{abstract}
In this work, we study the elastic scattering of some light particles, such as $^2$H, $^3$H, $^3$He and $^4$He, by heavy target nuclei 
with  an 
extended Watanabe model, which uses as input the neutron-nucleus and proton-nucleus optical potentials and the ground-state wave functions of 
the projectile. The nucleon-nucleus optical potential used in this work was obtained within a semi-microscopic nuclear matter approach, 
whose real 
and imaginary parts are provided by the first and second-order terms, respectively, of the Taylor expansion of the Brueckner-Hartree-Fock mass 
operator obtained with the reaction G-matrix built up with the Gogny force \cite{lopez21}. 
The angular distributions of the scattering of $^2$H, $^3$H, $^3$He, and $^4$He from different target nuclei and 
at a different incident energy of the projectile computed with this model are analyzed. 
The reaction cross sections corresponding to some   
of these scattering processes 
are also calculated. Our results are compared with the experimental values as well as with another 
Watanabe calculation where the nucleon-nucleus optical potential is provided by the phenomenological K\"oning-Delaroche model.
The limitations of the extended Watanabe model used in this work are also discussed.
\end{abstract}

\maketitle

\section{Introduction}

The theoretical analysis of the scattering of nucleons by nuclei is usually performed with the help of the optical model \cite{hodgson63}, 
which allows  to predict different observables such as the elastic scattering angular distributions, the total reaction cross sections, and the 
analyzing powers in a relatively simple way. 
The optical potential can be built up in two different manners. In the case of phenomenological optical potentials, one assumes some analytical 
profile for the radial dependence of the potential, usually of Woods-Saxon type, and fits their parameters to reproduce a selected set of 
measurements of different scattering observables in some particular reactions \cite{becchetti69,varner91,koning03}. The optical potential can also 
be determined from more microscopic grounds, which often are based on the fact that this potential could be identified with the mass-operator 
of the one-particle Green function \cite{bell59}.
Following this idea, Jeukene, Lejeune and Mahaux  derived a microscopic optical potential performing Brueckner-Hartree-Fock (BHF) calculations in  
nuclear matter using realistic nucleon-nucleon interactions. This microscopic optical potential was applied to finite nuclei using the Local 
Density Approximation (LDA) with parametrized nuclear densities \cite{jeukene74,jeukene76,jeukene77a, jeukene77b,bauge98}. 
However, effective forces such as the Skyrme \cite{vautherin72} or Gogny \cite{decharge80} interactions, which are specially designed to describe 
ground-state properties of finite nuclei have also been used to obtain a nucleon-nucleus microscopic optical potential
\cite{shen09,xu14,pilipenko10,pilipenko12,lopez21}.
The underlying idea, in this case, is that these forces can be regarded as effective parameterization of the $g$-matrix and used to obtain
perturbatively the first and second-order terms of the expansion of the  mass operator of the one-body Green function.

A different approximation to obtain the microscopic optical potential is the folding model 
\cite{sinha75,brieva77}, where the optical potential is computed by convolution of a complex two-body nucleon-nucleon effective interaction
\cite{arellano90,arellano07, aguayo08,amos00,khoa02,loan19} with the target nucleus density. It can be seen from basic folding formulas that 
the method generates the first-order term of the microscopic optical potential derived from the Feshbach's theory \cite{feshbach92}. 
This model is used quite often as a basis to describe the scattering of light projectiles by heavier nuclei. In this case, the corresponding microscopic optical
potential can be obtained by folding the projectile density with the nucleon-nucleus optical potential, which represents the interaction of a free 
nucleon of the projectile with the target nucleus \cite{satchler79,satchler83,brandon97,li09,pang11}.

In a previous paper \cite{lopez21} we have developed a microscopic  model to describe nucleon-nucleus scattering at relatively low bombarding 
energy using Gogny forces of the D1 family. The real and imaginary parts of this microscopic optical potential are obtained as the first and 
second-order terms of the Taylor expansion of the mass operator, respectively, which are calculated within the BHF method using the $g$-matrix 
built up with the effective Gogny force. This optical potential is applied to finite nuclei using the LDA with the neutron and proton densities 
computed within a quasi-local Hartree-Fock approximation with the same Gogny interaction \cite{soubbotin00,soubbotin03}. 
This nucleon-nucleus microscopic optical potential \cite{lopez21} does not contain free parameters to be 
adjusted to scattering data and gives a reasonably good agreement with the experimental results of differential cross sections and analyzing 
powers of neutron and proton elastic scattering by atomic nuclei along the whole periodic table. Very recently, this model has been 
used successfully to describe quasielastic proton-neutron charge exchange reactions \cite{lopez23}.

We want to continue analyzing the ability of this Gogny-based microscopic optical potential to describe other  scattering processes.
In particular, in this work, we perform an exploratory study of the elastic scattering
of light projectiles $^2$H, $^3$H, $^3$He and $^4$He by atomic nuclei through a relatively simple folding model.
Actually, we follow a similar strategy to that used in Refs.~ \cite{guo10,pilipenko15,guo14,guo09,guo11,kuprikov16}
for describing this type of reaction with a microscopic optical potential built up with effective Skyrme forces.
We are aware that this simple model corresponds to the free scattering of the nucleon of the projectile by the target nucleus disregarding 
other possible reaction channels, such as inelastic scattering and breakup and transfer reactions, which are taken on the average by the 
imaginary part of the optical potential. Thus the microscopic optical potential used in this work becomes an extended Watanabe model 
\cite{watanabe58} for describing the light particle-nucleus elastic scattering (the original Watanabe model takes into account only deuteron-nucleus 
elastic scattering). In order to account for the breakup effects, the more fundamental continuum discretized coupled channel (CDCC) method was 
introduced long ago by Rawitscher \cite{rawitscher74}. More recently, Mackintosh and Keeley have pointed out the relevance of the pickup reactions 
in the coupled channel calculations for a precise description of the $^2$He-nucleus \cite{keeley08}, $^3$He-nucleus \cite{mackintosh19} and 
$^3$H-nucleus \cite{keeley20,keeley23} elastic scattering. In order to check the reliability of our model, we have compared 
our results for some deuteron-induced reactions with the CDCC results reported in Ref.\cite{chau06}, using in both cases the phenomenoñogical 
optical potential of K\"oning-Delaroche as underlying nucleon-nucleus optical potential. Also we have estimated the impact of including the pickup 
channels on the elastic scattering calculation for two reactions by comparing our results with the predictions of the coupled reaction channel 
(CRC) calculations of Ref.~\cite{keeley08,keeley20}.  

The paper is organized as follows. In the first section, we summarize our theory. The second section is devoted to discussing the predictions of our 
model concerning several elastic scattering observables measured in different reactions induced by $^2$H, $^3$H, $^3$He, and $^4$He. Our conclusions 
are laid out in the last section. The Appendix summarizes the renormalization procedure of the Gogny-based nucleon-nucleus optical potentials. 

\section{Basic Theory}
     
To describe elastic scattering by light particles we use an extended version of the Watanabe model \cite{watanabe58}, 
which was initially devised for describing high energy scattering of deuterons by complex nuclei. 

\subsection{The Watanabe model}
In this model, it is assumed that the scattering of deuterons by heavier nuclei is described by the wave function solution of the following 
Schr\"odinger equation: 
\begin{eqnarray} 
&&\bigg\{-\frac{\hbar^2}{2M} {\nabla_{\bf R}}^2 - \frac{\hbar^2}{2\mu} {\nabla_{\bf s}}^2 + V_{12}({\bf s})\nonumber \\ 
&+& V_1\big({\bf R} + \frac{{\bf s}}{2}\big) + V_2\big({\bf R} - \frac{{\bf s}}{2}\big)\bigg\}\Psi({\bf{R}},{\bf{s}})
=E\Psi({\bf{R}},{\bf{s}}),
\label{eq1}
\end{eqnarray}
which is  written in terms of the neutron and proton center of mass and relative coordinates, ${\bf R}=({\bf r_1}+{\bf r_1})/2$ and 
${\bf s}={\bf r_1}-{\bf r_2}$. In this equation $M=m_n+m_p$ and $\mu=m_n m_p/(m_n+m_p)$ are the total and reduced mass of the two nucleons, $V_1$ 
and $V_2$ represents the interaction of the neutron and proton with the target nucleus and $V_{12}$ is the neutron-proton interaction in the 
deuteron.  

The wave function solution of the Schr\"odinger equation (\ref{eq1}) can be written as 
\cite{watanabe58}
\begin{equation}
\Psi({\bf{R}},{\bf{s}}) = \chi_0({\bf{s}})\Phi({\bf{R}}) + F({\bf{R}},{\bf{s}}), 
\label{eq2}
\end{equation}
where $\chi_0({\bf{s}})$ is the ground state-deuteron wave function, $\Phi({\bf{R}})$ describes the center of mass motion and $F({\bf{R}},{\bf{s}})$
 takes into account the coupling between the center of mass and the relative motion.

The ground-state deuteron wave function, which corresponds to total spin and isospin $S$=1 and $T$=0, respectively, does not contain $D$-wave 
admixture and is the solution of the intrinsic Schr\"odinger equation 
\begin{equation}
\bigg\{- \frac{\hbar^2}{2\mu} {\nabla_{\bf s}}^2 + V_{12}({\bf s})\bigg\} \chi_0 = \epsilon_d \chi_0,
\label{eq3}
\end{equation} 
which together with the wave functions $\chi_i (i=1,2,..)$, also solutions of the equation $H_{in} \chi_i = \epsilon_i \chi_i$ and orthogonal
to $\chi_0$, form a complete set of wave functions, which,  assuming a discretized continuum, allows to write 
\cite{chau06,kuprikov12,pilipenko15}
\begin{equation}
F({\bf{R}},{\bf{s}}) = \sum_{i>0} \chi_i({\bf{s}}) \phi_i({\bf{R}}).  
\label{eq4}
\end{equation}

Next, multiplying the Schr\"odinger equation (\ref{eq1}) by $\chi_0^{\dagger}$ and integrating over the relative coordinate one obtains
\begin{equation}
-\frac{\hbar^2}{2M} {\nabla_{\bf R}}^2 \Phi({\bf R}) + V({\bf R}) \Phi({\bf R}) +g({\bf R}) = (E - \epsilon_d) \Phi({\bf R}),
\label{eq5}
\end{equation}
where
\begin{equation}
V({\bf R}) = \int d{\bf s} \chi_0^{\dagger}({\bf s}) \bigg[V_1\big({\bf R} + \frac{{\bf s}}{2}\big) + V_2\big({\bf R} - \frac{{\bf s}}{2}\big)\bigg]
\chi_0({\bf s})
\label{eq6} 
\end{equation}
and
\begin{equation}
g({\bf R}) = \int d{\bf s} \chi_0^{\dagger}({\bf s}) \bigg[V_1\big({\bf R} + \frac{{\bf s}}{2}\big) + V_2\big({\bf R} - \frac{{\bf s}}{2}\big)\bigg]
F({\bf{R}},{\bf{s}}).
\label{eq7}
\end{equation}
The last contribution in the left-hand side of (\ref{eq5}) $g({\bf R})$ contains the breakup of the deuteron and the distortion of the deuteron wave
 function. These effects in the simplest version of the Watanabe model are approximated by adding an imaginary contribution to $V({\bf R})$ and 
dropping $g({\bf R})$. In a last step, the potentials $V_1$ and $V_2$ in (\ref{eq7}) are replaced by the corresponding neutron-nucleus and 
proton-nucleus optical potentials.
The dipole polarizability in the deuteron scattering is a manifestation of the coupling to breakup channels via the electric dipole 
operator \cite{andres94}. This effect has an important effect on energies around the Coulomb barrier \cite{moro99}. It was pointed out some time ago
 that the trivially local equivalent polarization potential describing the effect of the breakup channels on elastic scattering has both real and 
imaginary contributions  \cite{austern87}. 
Although the breakup effects and the impact of the dipole polarizability have not been included explicitly in our model, we expect they are included,
 at least partially, in an effective way through the imaginary part of the optical potential, as in the original Watanabe work \cite{watanabe58}. 
We have checked our approach by performing some comparisons between the differential and total cross sections computed with the original Watanabe 
 model and the CDCC results of Ref.\cite{chau06}. 
  
\subsection{The extended Watanabe model}
In order to describe the elastic scattering of $^3$H, $^3$He, and $^4$He by heavier nuclei, we use a 
generalized Watanabe model where the microscopic projectile-target optical potential is obtained by folding the nucleon-target optical potential 
of the constituents nucleons in the ground-state of the projectile. It can be written as: 
\begin{equation}
V({\bf R}) = \sum_{i=1}^n \langle \chi_0({\bf \xi}_1,...{\bf \xi}_{n-1})\vert V_i({\bf R},{\bf \xi}_1...{\bf \xi}_{n-1})\vert
\chi_0({\bf \xi}_1,...{\bf \xi}_{n-1}) \rangle,
\label{potential}
\end{equation}
where $n$ is the number of nucleons in the projectile, $V_i({\bf R},{\bf \xi}_1...{\bf \xi}_{n-1})$ is the free nucleon-target optical
potential corresponding to the nucleon $i$ and $\chi_0({\bf \xi}_1,...{\bf \xi}_{n-1})$ is the ground-state wave function of the projectile, both 
written in terms of the Jacobi coordinates. In this system ${\bf R}$ is the center of mass position of the projectile and ${\bf \xi}_i (i=1...n-1)$
 are the relative coordinates related to the positions ${\bf r}_1,{\bf r}_2......{\bf r}_n$ of each nucleon in the projectile by:
\begin{eqnarray}
&&{\bf R} = \frac{1}{n} \sum_{i=1}^n {\bf r}_i \nonumber \\
&&{\bf \xi}_j = \frac{1}{j} \sum_{k=1}^j {\bf r}_k - {\bf r}_{j+1} \quad (j=1,2...n-1),
\label{coordinates}
\end{eqnarray}
where $n$=2 for $^2$H, $n$=3 for $^3$H and $^3$He and $n$=4 for $^4$He. 

The wave functions of the different projectiles, which enter in (\ref{potential}), are expressed  in terms of the intrinsic coordinates. In the 
case of the deuteron the ground-state wave function, assuming only $s$-wave contribution, is expressed by the Hulth\'en 
function \cite{guo10} 
\begin{equation}
\Phi_{^2H}(\xi_1) = \frac{N_d}{\xi_1}\big[e^{-\alpha \xi_1}-e^{-\beta \xi_1}\big],
\label{wfA2}
\end{equation}
where $N_d = \sqrt{\frac{\alpha \beta(\alpha+\beta)}{2\pi(\beta-\alpha)^2}}$ with $\alpha$=0.23 fm$^{-1}$ and $\beta$=1.61 fm$^{-1}$.

For $^3$H and $^3$He the spatial part of the intrinsic wave functions are expressed by the three-dimensional harmonic oscillator functions as  
\cite{guo14,guo09}
\begin{equation}
\Phi_{^3H}(\xi_1,\xi_2)=\Phi_{^3He}(\xi_1,\xi_2)= \bigg(\frac{\beta^2}{3\pi^2}\bigg)^{\frac{3}{4}} e^{-\frac{\beta\xi_1^2}{4}-
\frac{\beta\xi_2^2}{3}},
\label{wfA3}
\end{equation}
where $\beta$=0.346 fm$^{-2}$ for $^3$H and $\beta$=0.283 fm$^{-2}$ for $^3$He. Notice that the full wave functions of these two nuclei also 
contain, in addition to the spatial part, spin, and isospin contributions. In order to preserve the antisymmetry, the two identical particles are 
coupled to isospin $T$=1 and spin $S=0$. When the third nucleon is added, the total spin and isospin become $S_t=1/2$ and $T_t$=1/2, because the 
ground-states of $^3$H and $^3$He are a isospin doublet. These total spin and isospin values together with the symmetric spatial part of the wave 
functions of these nuclei given by Eq.(\ref{wfA3}) implies that in their ground-state they have spin-parity $J^{\pi}$=1/2$^{+}$, which is in 
agreement with the experimental values. However, as far as the optical potential (\ref{potential}) only depends on spatial variables, the spin and 
isospin structure of the $^3$H and $^3$He do not play any role in its calculation.

Again using harmonic oscillator wave functions the wave function of $^4$He can be written as
\begin{equation}
\Phi_{^4He}(\xi_1,\xi_2,\xi_3)=\bigg(\frac{\beta^3}{4\pi^3}\bigg)^{\frac{3}{4}} e^{-\frac{\beta(\xi_1^2+\xi_2^2)}{4}-\frac{\beta\xi_3^2}{2}},
\label{wfA4}
\end{equation}
where in this case $\beta$=0.4395 fm$^{-2}$.
 
\subsection{The nucleon-nucleus optical potential}
The other piece of the Watanabe modeĺ is the free nucleon-nucleus optical potential, which we have derived in a previous publication \cite{lopez21} 
on the basis of a semi-microscopic nuclear matter approximation using Gogny forces of the D1 family. Gogny interactions were introduced by D. Gogny 
in the early eighties \cite{decharge80} aimed to describe simultaneously the mean field and the pairing field of finite nuclei with the same 
interaction. The Gogny D1S parameterization \cite{berger91} has been used in large-scale Hartree-Fock-Bogoliubov calculations of ground-state 
properties of finite nuclei along the whole periodic table \cite{cea}. A detailed analysis of these results shows some deficiencies in the 
theoretical description of masses of neutron-rich nuclei compared with the corresponding experimental values (see \cite{pillet17} for more details).
 To remedy these limitations of D1S, new parameterizations of the Gogny force, namely D1N \cite{chappert08} and D1M \cite{goriely09}, have been 
proposed. These forces incorporate in their fitting protocol the constraint of reproducing, in a qualitative way, the microscopic Equation of State
 of Friedman and Pandharipande in neutron matter in order to improve the description of neutron-rich nuclei.

The Gogny forces of the D1 family consist of a finite-range part and a zero-range density-dependent term together with spin-orbit interaction, 
which is also zero-range  as in the case of Skyrme forces. The finite-range part is the sum of two Gaussian form factors with different ranges, each 
multiplied by all the possible spin-isospin exchange operators with different weights. Therefore this type of Gogny force read:
\begin{eqnarray}
&&V(\vec{r}_{12})=t_3(1+\hat{P}_\sigma)\delta(\vec{r}_{12})\left[\rho\left(\frac{\vec{r}_1+\vec{r}_2}{2}\right)\right]^{1/3}\nonumber \\
&+&\sum_{k=1}^{k=2}e^{-\left(\frac{\vec{r}_{12}}{\mu_k}\right)^2} 
\times \left(W_k+B_k\hat{P}_\sigma-H_k\hat{P}_\tau-M_k\hat{P}_\sigma\hat{P}_\tau\right)\nonumber \\
&+&
iW_{LS}\left(\hat{\sigma}_1+\hat{\sigma}_2\right)\cdot\hat{k}^{\dag}\times\delta\left(\vec{r}_{12}\right)\hat{k} \label{eq9}
\end{eqnarray}
where
\[\hat{P}_\sigma=\frac{1}{2}\left(1+\hat{\sigma}_1\cdot\hat{\sigma}_2\right) \quad \textrm{and} \quad\hat{P}_\tau=\frac{1}{2}\left(1+\hat{\tau}_1
\cdot\hat{\tau}_2\right)\]
\noindent
are the spin and isospin exchange operators, respectively, while \\
\[\vec{r}_{12}=\vec{r}_1-\vec{r}_2,\quad \textrm{and} \quad\hat{k}=\frac{1}{2i}\left(\vec{\nabla}_1-\vec{\nabla}_2\right)\]
\noindent
are the relative coordinate and the relative momentum of the two nucleons, respectively. The parameters of the force, namely $W_k, B_k, H_k, M_k,
\mu_k$ ($k$ =1, 2), $t_3$, and $W_{LS}$ are fitted to reproduce some properties of finite nuclei and infinite nuclear matter (see Refs.
\cite{decharge80} and \cite{goriely09} for more details about the fitting protocol of Gogny interactions). \\

Let us summarize the more relevant aspects of the microscopic optical potential based on the Gogny interaction, which was derived in 
Ref.\cite{lopez21} and will be used in the present work. It is obtained with the Jeukene, Lejeune and Mahaux procedure 
\cite{jeukene74,jeukene76,jeukene77a,jeukene77b} in 
nuclear matter. The real and imaginary parts of the central potential are determined from the first and second terms of the Taylor expansion of 
the mass operator, which is computed using the effective 
Gogny force instead of a microscopic interaction.  In the present work, we use the D1S parameterization of the Gogny force, whose parameters 
are given in Table \ref{parameters}. In Ref.\cite{lopez21} we checked that the use of other parameterizations of the Gogny force such as D1N 
or D1M provides basically the same description of the nucleon-nucleus elastic scattering. The model described here based on the Gogny interaction 
is similar to other microscopic optical potentials obtained from the same theoretical grounds but using Skyrme forces 
\cite{shen81,shen09,xu14,pilipenko10,pilipenko12}.

\begin{table}[ht]
\begin{center}
\caption{Parameters of the effective D1S Gogny force used in this work.}
\begin{tabular}{|l|l|l|l|l|l|l|} \hline & $k$ &$\mu_k(fm)$&$W_k$&$B_k$&$H_k$&$M_k[MeV]$
\\ \hline D1S &1 &0.7&-1720.3&1300&-1813.53&1397.60 \\ & 2&1.2 &103.64&-163.48&162.81&-223.93
\\ \hline
\end{tabular}
\label{parameters}
\end{center}
\begin{center}
\begin{tabular}{|l|l|l|}\hline D1S & $W_{LS}=130[MeV\cdot fm^5]$ & $t_3=1390[MeV\cdot fm^4]$
\\ \hline
\end{tabular}
\end{center}
\end{table}

In this model, the real part of the central potential corresponds to the single-particle potential felt by the incident projectile due to the nucleon
 in the target. In the strict application of the model, this real part is the Hartree-Fock (HF) single-particle potential in nuclear matter 
considering only the two-body part of the nucleon-nucleon interaction, which for an incident nucleon $\alpha$ of type $\tau$ reads:
\begin{widetext}
\begin{eqnarray}
V_{\tau\alpha}=\frac{3}{2}t_3\rho^{1/3}[\rho-\rho_{\tau\alpha}]+\pi^{3/2}\sum_{k=1}^{k=2}\mu_k^3\Bigg[\left(W_k+\frac{B_k}{2}\right)\rho-
\left(H_k+\frac{M_k}{2}\right)\rho_{\tau\alpha}\Bigg]-\nonumber\frac{1}{4\pi^{3/2}}\times\\ \sum_{k=1}^{k=2}\mu_k^3\Bigg[\left(\frac{W_k}{2}
+B_k-\frac{H_k}{2}-M_k\right)I(k_\alpha,k_\rho=k_{\tau\alpha})-\left(\frac{H_k}{2}+M_k\right)I(k_\alpha,k_\rho=k_{-\tau\alpha})\Bigg],\nonumber \\
\label{eq12}
\end{eqnarray}
\end{widetext}
where the Fermi momenta of the particles with the same (opposite) isospin as the projectile are related to the corresponding particle densities by
 $k_{\tau\alpha}^3=3\pi^2\rho_{\tau\alpha}$ ($k_{-\tau\alpha}^3=3\pi^2\rho_{-\tau\alpha}$). In Eq.(\ref{eq12}) the functions $I$, which depend on 
the momentum $k_{\alpha}$ of the projectile and correspond to the HF exchange potential, are defined as
\begin{widetext}
\begin{eqnarray}
I(k_\alpha,k_\rho)=\frac{4\pi^{3/2}}{\mu_k^3}\left[erf\left(\frac{\mu_k}{2}(k_\rho-k_\alpha)\right)
+ erf\left(\frac{\mu_k}{2}(k_\rho+k_\alpha)\right)\right]
+\frac{8\pi}{k_\alpha\mu_k^4}\left[e^{-\frac{\mu_k^2}{4}(k_\alpha+k_\rho)^2}-e^{-\frac{\mu_k^2}{4}(k_\alpha-k_\rho)^2}\right].
\label{eq13}
\end{eqnarray} 
\end{widetext}
To obtain the real part of the central potential in a finite nucleus we apply the LDA in (\ref{eq12}) with the self-consistent neutron 
and proton densities of 
the target computed within the quasi-local density formalism \cite{soubbotin00,soubbotin03,lopez21}, which allows to express $V_{\tau\alpha}$ as 
a function of the position through the densities and local Fermi momenta of neutrons and protons in Eq.(\ref{eq12}). The radial dependence of the 
momentum of the incident neutron $k_{\alpha}$ is given by the solution of the equation:
\begin{equation}
E_L = \frac{\hbar^2 k^2_{\alpha}}{2m} + V_{\tau\alpha}(k_{\alpha},k_{\tau\alpha}(R),k_{-\tau\alpha}(R)) \label{eq14}
\end{equation}
where $E_L$ is the energy of the projectile in the laboratory frame (if the projectile is a proton the Coulomb potential of the target has to be 
added to (\ref{eq14})). Due to the finite range of the Gogny force the momentum of the incident nucleon $k_{\alpha}$ appears in the kinetic and 
potential contributions to Eq.(\ref{eq14}), which implies that this equation has not an exact analytical solution and therefore the momentum  
$k_{\alpha}$ must be obtained numerically. However, the momentum of the incident nucleon can still be obtained approximately in an analytical way
 by performing locally a quadratic Taylor expansion of $V_{\tau\alpha}$ around the Fermi momentum  $k_{\tau\alpha}$ of the particles with the same 
isospin as the projectile, i.e.
\begin{eqnarray}
&&V_{\tau\alpha}(R)=V_{\tau\alpha}(k_{\alpha}=k_{\tau\alpha},R)\nonumber \\
&+& \left[\frac{1}{2k_{\alpha}}\frac{\partial V_{\tau\alpha}(k_{\alpha},R)}
{\partial k_{\alpha}}\right]_{k_{\alpha}=k_{\tau\alpha}}(k_{\alpha}^2-k_{\tau\alpha}^2), \label{eq15}
\end{eqnarray}
where the coordinate $R$ indicates the radial dependence of the real part of the central potential, owing to its dependence on the 
local neutron and proton Fermi momenta of the target. With this approximation Eq.(\ref{eq14}) can be recast as
\begin{equation}
E_L = \frac{\hbar^2 k^2_{\alpha}}{2m^*_{\tau\alpha}} + V_{0\tau\alpha}(R), \label{eq16}
\end{equation}
where $m^*_{\tau\alpha}$ is the effective mass at the Fermi momentum $k_{\tau\alpha}$, defined as
\begin{equation}
\frac{m}{m^*_{\tau\alpha}} = 1 + \frac{m}{\hbar^2}\frac{1}{k_{\tau\alpha}}\left[\frac{\partial V_{\tau\alpha}}{\partial k_{\alpha}}\right]
_{k_\alpha=k_{\tau\alpha}} \label{eq17}
\end{equation}
and 
\begin{equation}
V_{0\tau\alpha}=V_{\tau\alpha}(k_{\alpha}=k_{\tau\alpha})-\frac{k_{\tau\alpha}}{2}\left[\frac{\partial V_{\tau\alpha}}{\partial k_\alpha}.
\right]_{k_\alpha=k_{\tau\alpha}} \label{eq18}
\end{equation}

The imaginary part of the central potential is given by \cite{lopez21}
\begin{widetext}
\begin{eqnarray}
W_{\alpha}&=&\frac{1}{2}Im \sum_{\nu\leq k_F \atop \lambda,\mu > k_F}<\alpha\nu|V|\widetilde{\lambda\mu}>
\frac{1}{\epsilon_{\alpha}+\epsilon_{\nu}-\epsilon_{\lambda}-\epsilon_{\mu}.
+i\eta}<\lambda\mu|V|\widetilde{\alpha\nu}> \label{eq19}
\end{eqnarray}
\end{widetext}
The denominator of this expression can be easily worked out by making use of the parabolic approach, which allows writing the single particle energy 
of the projectile as (\ref{eq16}) and similar expressions for the other particle and hole states entering in Eq.(\ref{eq19}). Due to the charge 
conservation, we can write the denominator of (\ref{eq19}) as

\begin{equation}
\varepsilon_{\alpha} + \varepsilon_{\nu} - \varepsilon_{\lambda} - \varepsilon_{\mu} 
=\frac{\hbar^2k^2_{\alpha}}{2m^*_{\tau\alpha}} + \frac{\hbar^2k^2_{\nu}}{2m^*_{\tau\nu}} 
- \frac{\hbar^2k^2_{\lambda}}{2m^*_{\tau\lambda}} - \frac{\hbar^2k^2_{\mu}}{2m^*_{\tau\mu}}. 
\label{eq20}
\end{equation}
Applying the principal value integral to deal with the denominator of (\ref{eq19}), after some lengthy algebra one obtains the imaginary part of 
the central potential in the nuclear matter approach, which can be finally expressed as 
\begin{equation}
W_{\tau\alpha}=-\frac{1}{2}\frac{\pi}{(2\pi)^6}\left[W_1+2W_2+W_3+W_4+W_5\right] \label{eq21}
\end{equation}
where the different contributions to (\ref{eq21}) are given in \cite{lopez21}.

The microscopic nucleon-nucleus optical potential that describes the light-particle target scattering (\ref{potential}), also contains 
Coulomb and spin-orbit contributions. The Coulomb potential in the light-particle microscopic optical potential is included in the different 
proton-nucleus potentials entering in Eq.(\ref{potential}). Except in the case of $^4$He where it vanishes, the spin-orbit potentials in the other 
light particles considered
 in this work are consistent with the sum of the spin-orbit potentials for its constituent nucleons. Notice however that the spin-orbit for the 
deuteron is multiplied by an energy factor \cite{guo10}, which takes into account the energy dependence of the spin-orbit potential in this case, and
 for tritium the spin-orbit energy of each constituent is multiplied by the corresponding effective mass in order to enhance the spin-orbit 
contribution at the nuclear surface \cite{guo14}. Let us finally mention that in the calculations reported in the next section, we have replaced  
the real part of the single-particle potential in the nuclear matter approach with the one obtained in the HF approximation, which takes into 
account the finite size of the target nucleus as well as the rearrangement contributions (see \cite{lopez21} for more details).   
   
\section{Results}
\subsection{Optical potentials for the scattering of light projectiles from nuclei}
As we have discussed previously, the microscopic optical potential for describing the elastic scattering of deuterons from nuclei used in this work 
is built up with the help of the Watanabe model and is given by Eq.(\ref{eq6}), where $V_1$ and $V_2$ are the neutron-nucleus and proton-nucleus 
microscopical optical potentials based on the Gogny D1S effective force and derived in Ref.~\cite{lopez21}. 

Using the optical potential provided by the extended Watanabe model (\ref{potential}) with the intrinsic wave functions for $^3$H, $^3$He and 
$^4$He given by Eqs.(\ref{wfA3}) and (\ref{wfA4}), the microscopic optical potential for describing $^3$H and $^3$He scattering from heavier nuclei 
reads
\begin{widetext}
\begin{eqnarray}
V_{^3H(^3He)}({\bf R}) = \int d{\bf \xi_1} d{\bf \xi_2}  \chi_0^{\dagger}({\bf \xi_1,\xi_2})
\bigg[V_{n(p)} \big({\bf R} + \frac{{\bf \xi_1}}{2} + \frac{{\bf \xi_2}}{3}\big) +
V_{n(p)} \big({\bf R} - \frac{{\bf \xi_1}}{2} + \frac{{\bf \xi_2}}{3}\big) 
+ V_{p(n)} \big({\bf R} - \frac{2{\bf \xi_2}}{3}\big)\bigg] \chi_0({\bf \xi_1,\xi_2}),
\label{eq222}
\end{eqnarray}
\end{widetext}
and for scattering of $\alpha$ particles is given by
\begin{widetext}
\begin{eqnarray}
V_{^4He)}({\bf R}) &=& \int d{\bf \xi_1} d{\bf \xi_2} d{\bf \xi_3}  \chi_0^{\dagger}({\bf \xi_1,\xi_2,\xi_3})
\bigg[V_{n} \big({\bf R} + \frac{{\bf \xi_1}}{2} + \frac{{\bf \xi_3}}{2}\big) +
V_{n} \big({\bf R} - \frac{{\bf \xi_1}}{2} + \frac{{\bf \xi_3}}{2}\big) 
\nonumber \\
&+& V_{p} \big({\bf R} + \frac{{\bf \xi_2}}{2} - \frac{{\bf \xi_3}}{2}\big)
+  V_{p} \big({\bf R} - \frac{{\bf \xi_2}}{2} - \frac{{\bf \xi_3}}{2}\big)\bigg] \chi_0({\bf \xi_1,\xi_2,\xi_2)},
\label{eq223}
\end{eqnarray}
\end{widetext}
where $V_n$ and $V_p$ are the neutron-nucleus and proton-nucleus optical potentials  obtained as explained in Section 2.3. At this point, two comments are in order. First, within the theoretical model used in this work, the energy $E$ of the projectile is shared among its 
nucleons, which means that the neutron-nucleus and proton-nucleus optical potentials entering in Eqs. (\ref{eq6}), (\ref{eq222}) and (\ref{eq223}) 
are computed at energy values 
of $E/2$, $E/3$ and $E/4$, respectively. Second, if this energy is smaller than the Coulomb barrier of the proton-target system, the real and 
imaginary parts of the central potential are not taken into account and the proton-nucleus optical potential reduces to its Coulomb 
contribution. However, this prescription may  be an oversimplification that underestimates the reaction cross-section because, as it is explained in 
Ref.\cite{lei19}, in the 
scattering of a composite weakly bound particle, such as $^2$H, $^3$H or $^3$He, there is a reaction probability even in the case that the energy 
of a fragment is below that of the fragment-target Coulomb barrier due to the Trojan horse effect \cite{baur76}.
\begin{figure}[ht]
\centering
\includegraphics[width=1.\linewidth]{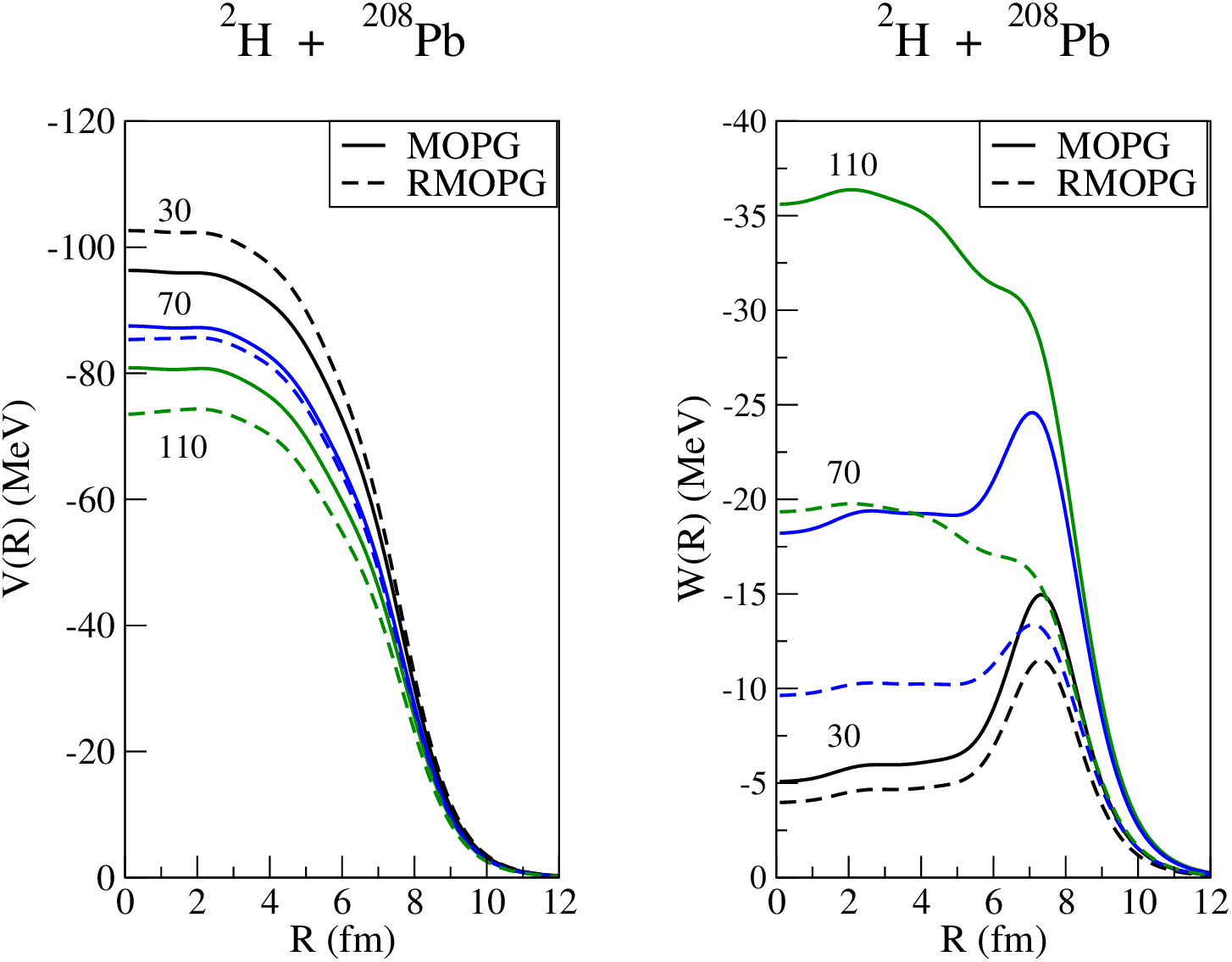} \\
\vspace{1cm}
\includegraphics[width=1.\linewidth]{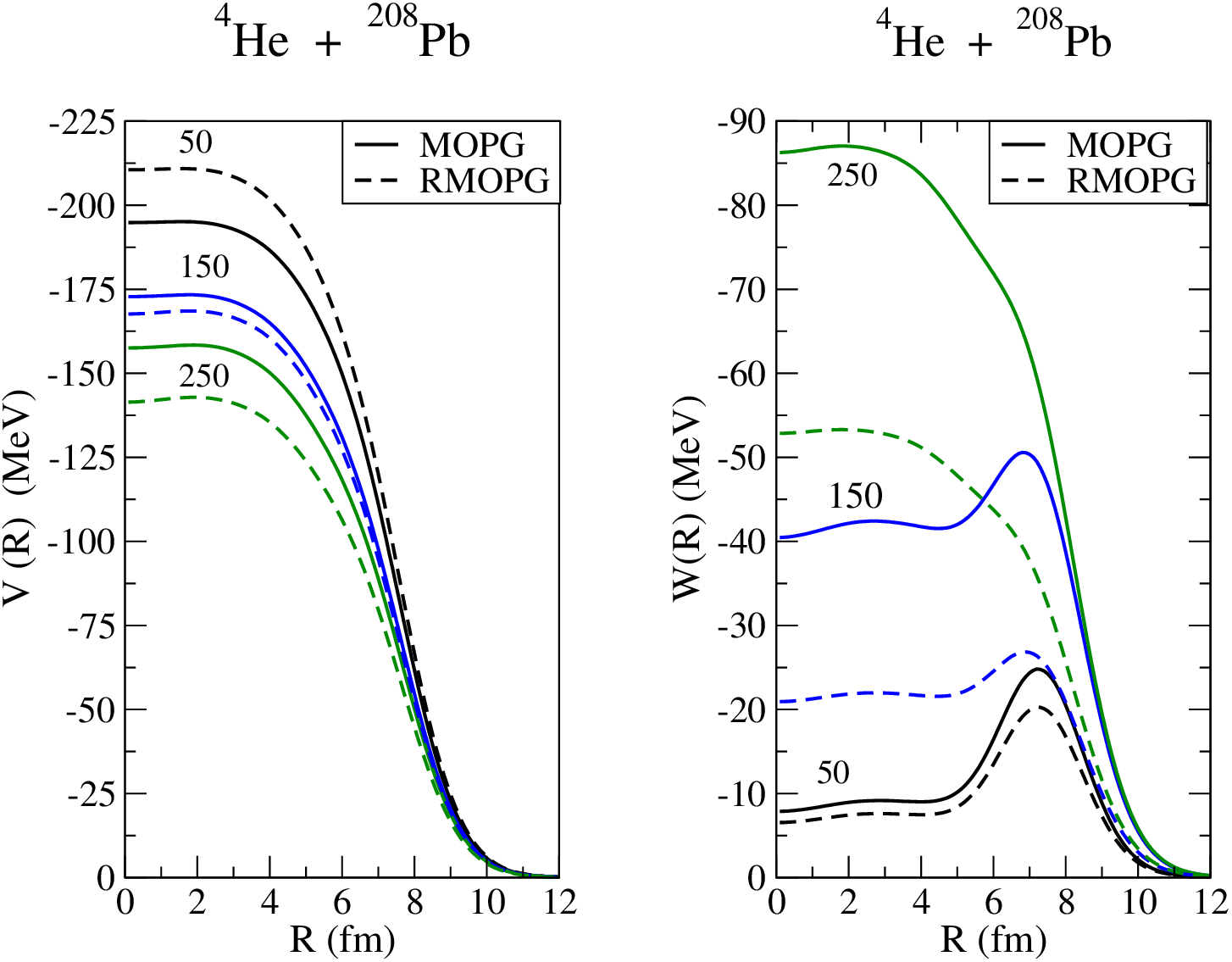}
\caption{Real and imaginary parts of the $^2$H-$^{208}$Pb (up) and $^4$He-$^{208}$Pb (down) microscopic optical potential computed with our 
microscopic unrenormalized (MOPG) and renormalized (RMOPG) models at several energies of the projectile.}
\label{fig1}
\end{figure}
\begin{figure}[ht]
\centering
\includegraphics[width=1.\linewidth]{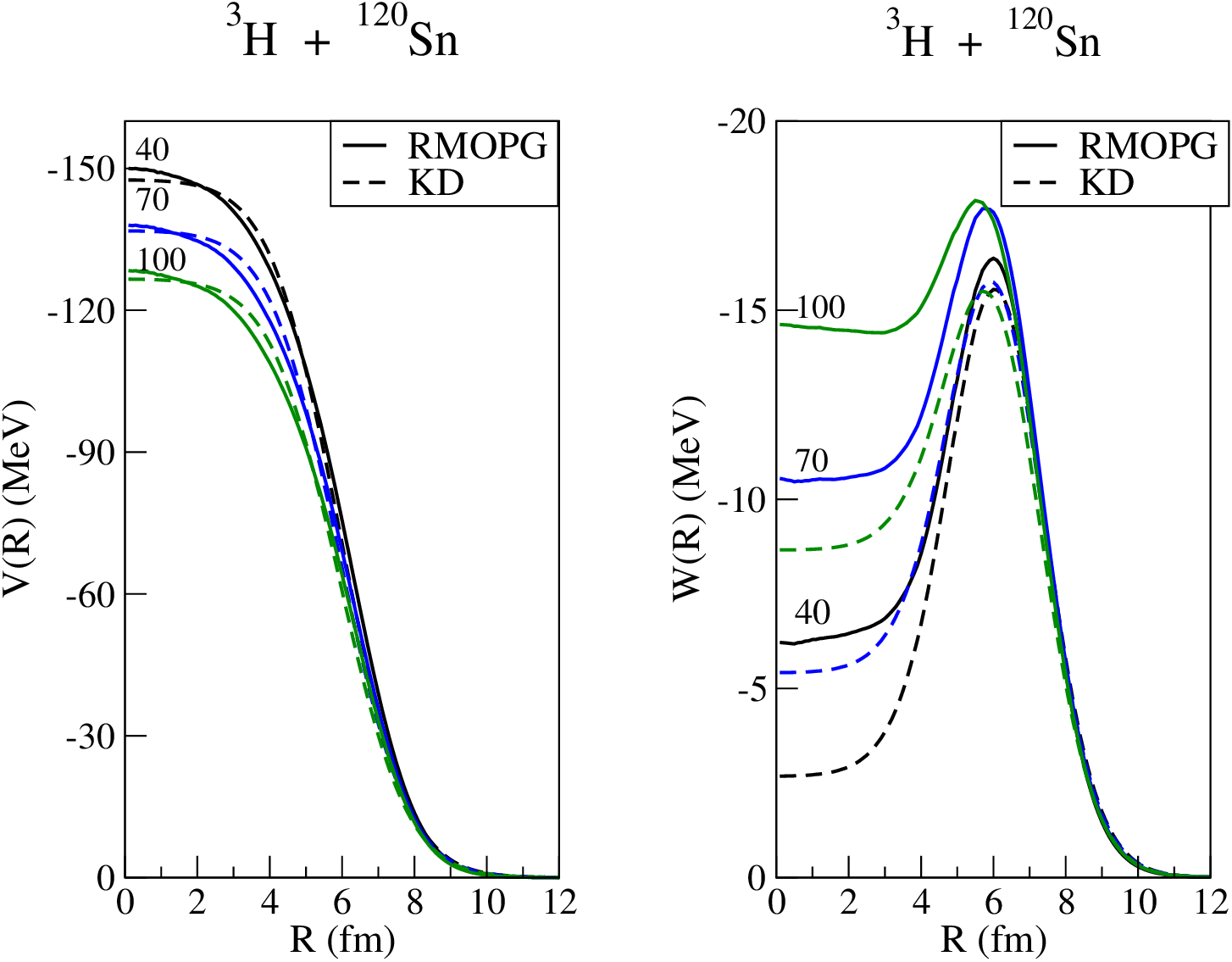} \\
\vspace{1cm}
\includegraphics[width=1.\linewidth]{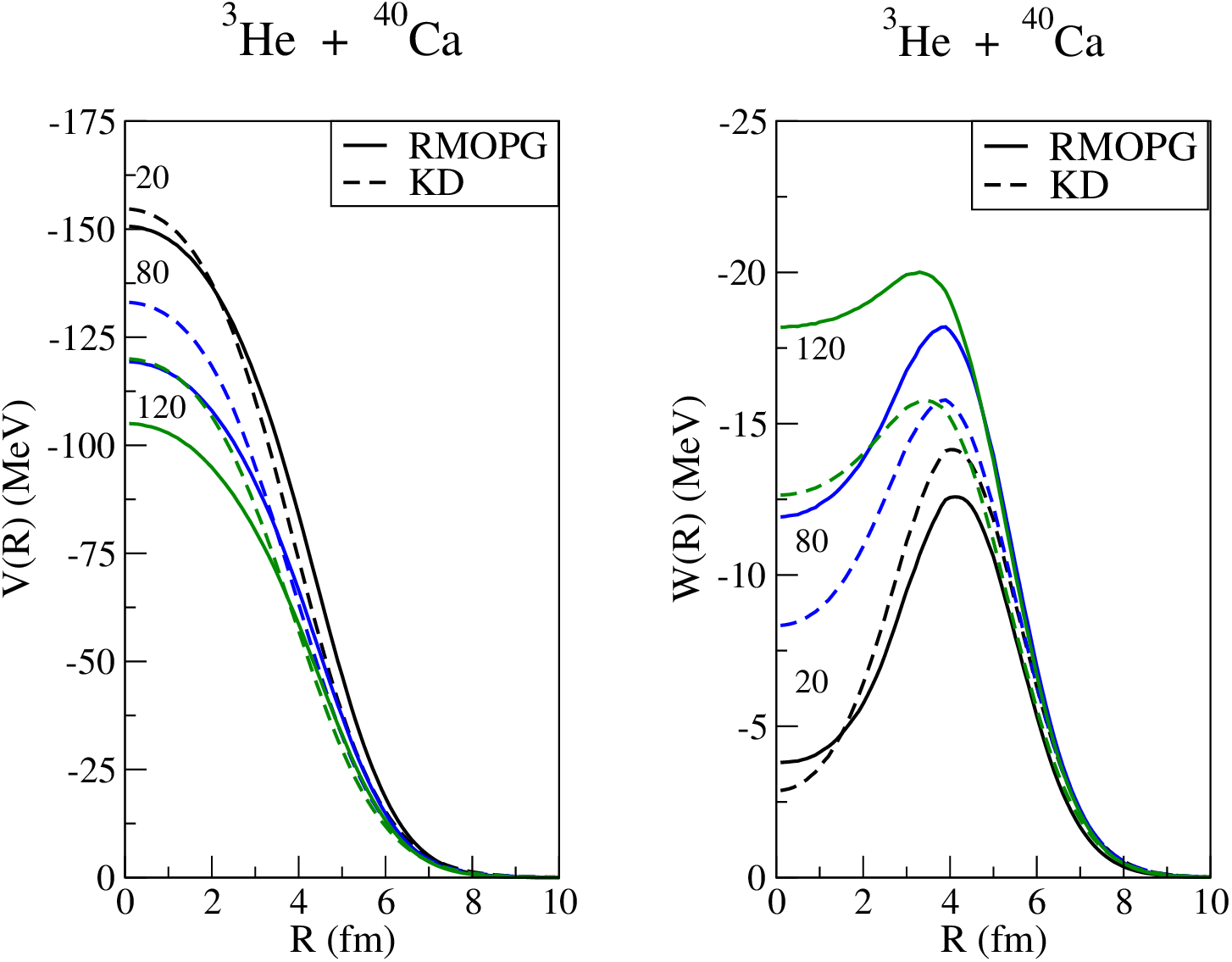}
\caption{Real and imaginary parts of $^3$H-$^{120}$Sn (up) and $^3$He-$^{40}$Ca (down) microscopic optical potentials computed with our 
renormalized model (RMOPG) and with a Watanabe model based on the K\"oning-Delaroche (KD) nucleon-nucleus optical potential computed at several 
energies of the projectile.}
\label{fig1a}
\end{figure}

The real and imaginary parts of the central contribution to the microscopic optical potential 
 based on the Gogny force for describing the elastic scattering of light particles 
from heavier targets (MOPG therein after), show, as a function of the energy of the incident projectile, some global trends that are largely 
independent of the projectile and of the target nucleus. Just as an example, we show in the different panels of Figure \ref{fig1} the evolution 
with the energy of the projectile of the profiles of real and imaginary parts of the central term of the MOPG provided by Eqs.(\ref{eq6}) and 
(\ref{eq223}) for the reactions $^2$H-$^{208}$Pb and  $^4$He-$^{208}$Pb. From this Figure, we can see that in all the cases the depth of the real 
part decreases with increasing energy, which points out the repulsive character of the energy dependence of the real part of the MOPG. However, 
the opposite trend happens for the imaginary part and we see that its strength increases when the energy of the projectile grows, due to the fact 
that at higher energy more inelastic channels are open. On the other hand, the imaginary part computed with the MOPG develops, for both reactions, 
a very well-marked peak at the surface, which indicates its strong absorptive character at the surface of the target. When the energy of the 
projectile increases, the volume absorption in the interior of the target is more relevant, as far as the higher energy of the projectile allows to
 explore  the inner part of the target. This volume absorption also increases with the mass of the target and becomes stronger than the surface 
absorption in the case of heavy $^{208}$Pb targets. These trends are also observed in other similar calculations of the optical
potential  designed for describing the
 scattering of light particles based on the Skyrme forces \cite{guo10,guo14,guo09,guo11}. It is interesting to point out that the central real and 
imaginary parts of the phenomenological optical potentials for describing the elastic scattering of $^2$H \cite{han06}, $^3$H and $^3$He 
\cite{pang09} and $^4$He \cite{su15} also show for a given target similar trends as a function of the energy.

As it has been pointed out in our previous work \cite{lopez21}, the nucleon-nucleus microscopic optical potential based on the Gogny
 force predicts more absorption, in particular at high energy of the projectile, but roughly similar real part than phenomenological optical 
potentials, as for example 
K\"oning-Delaroche (see in this respect Figures 1 and 2 of \cite{lopez21}). We expect that this behavior of the nucleon-nucleus potential may 
have a relevant impact on the MOPG owing to Eq. (\ref{potential}). In order to get an improvement of the description of the elastic 
scattering of light particles by nuclei using the MOPG model, inspired by Ref.\cite{bauge98} we have renormalized the central real and imaginary 
parts and the spin-orbit term of the Gogny nucleon-nucleus optical potential by energy-dependent factors following the protocol discussed in the 
Appendix. This renormalized MOPG (RMOPG from now on) is also displayed in Figure \ref{fig1}. The renormalization makes the real part deeper at low 
bombarding energies and shallower at high energies following, roughly, a smooth transition. As a function of the energy of the projectile, the 
renormalization reduces strongly the imaginary part at high energy but only a little at low energy. Also, the surface bump is damped by the 
renormalization. The RMOPG predictions are quite similar to those obtained by using an extended Watanabe potential Eq.(\ref{potential}) built up 
using the phenomenological nucleon-nucleus optical potential of K\"oning and Delaroche (KD in the following), as it can be seen in Figure 
\ref{fig1a} where we display the real and imaginary parts for the $^3$H + $^{120}$Sn and $^3$He +$^{40}$Ca reactions at several energies computed 
with both  models. For the first reaction, the real part predicted by the RMOPG and KD  models are almost identical while for the second reaction, the
 predictions of these two models differ more between them, probably due to the smaller mass of the target. Regarding the imaginary central part, 
the RMOPG and KD models predict a surface bump, which for the  $^3$H + $^{120}$Sn is roughly independent of the energy whereas for the 
$^3$He +$^{40}$Ca shows an increasing trend with increasing energy of the incident particle, this behavior being more pronounced in the RMOPG case.
 The imaginary part also contains a volume absorption region, which increases with the growing energy of the projectile and with the mass of the 
target in both models.

\begin{table}[ht]
\begin{center}
\caption{Real and imaginary parts of the phenomenological optical potentials (POP) that describe the scattering of $^2$H \cite{han06}, $^3$H and 
$^3$He \cite{pang09} and $^4$He at the 
$R_{SA}$ for different reactions at several energies. The predictions of the theoretical models MOPG, RMOPG, and KD at the phenomenological $R_{SA}$
 are also given.}
\scalebox{0.80}{
\begin{tabular}{cccccc}
\cline{1-6}
\hline
E & $R_{SA}$(fm) & $V_{POP}$(MeV) & $V_{KD}$(MeV) & $V_{MOPG}$(MeV) & $V_{RMOPG}$(MeV) \\
  & & $W_{POP}$(MeV) & $W_{KD}$(MeV) & $W_{MOPG}$(MeV) & $W_{RMOPG}$(MeV) \\
\hline
\multicolumn{6}{c}{$^{2}$H + $^{208}$Pb}\\
\hline
30  & 10.772 & -0.618 & -1.625 & -1.259 & -1.339   \\
    &        & -1.412 &  -0.732 &  -0.603 &  -0.461   \\
70  & 10.384 & -0.856 &  -2.189 &  -1.866 &  -1.809   \\
    &        & -2.321 &  -1.084 &  -1.725 &  -0.967   \\     
110 & 10.240 & -0.812 &  -2.188 &   -2.087 &  -1.880  \\
    &        & -2.875 &  -1.116 &  -2.266 &  -1.246   \\  
\hline
\multicolumn{6}{c}{$^{3}$H + $^{120}$Sn}\\
\hline
40  &  9.536 & -1.394 & -1.838 & -1.266 & -1.366   \\
    &        & -1.308 & -0.803 & -0.785 & -0.639   \\
70  &  9.140 & -2.167 & -2.924 & -2.326 & -2.414   \\
    &        & -1.487 & -1.386 & -2.009 & -1.296   \\
100 &  8.872 & -2.876 & -3.862 & -3.410 & -3.433   \\
    &        & -2.279 & -1.826 & -3.343 & -1.827   \\
\hline
\multicolumn{6}{c}{$^{3}$He + $^{40}$Ca}\\
\hline
20  &  8.002 & -0.848 & -1.305  & -0.863  & -0.972   \\
    &        & -1.001 &  -0.535 &  -0.359 & -0.359   \\
80  &  7.040 & -2.514 &  -3.995 &  -3.841 & -3.848   \\
    &        & -2.401 &  -2.086 &  -3.493 & -2.130   \\
120 &  6.729 & -3.464 &  -5.252 &  -5.683 & -5.369   \\
    &        & -3.004 &  -2.642 &  -5.725 & -3.015   \\
\hline
\multicolumn{6}{c}{$^{4}$He + $^{208}$Pb}\\
\hline
50  & 10.922 & -1.607 & -1.958 & -1.234 & -1.330   \\
    &        & -0.825 & -0.965 & -0.440 & -0.365   \\
150 & 10.217 & -3.040 & -4.348 & -3.619 & -3.459   \\
    &        & -2.478 & -2.469 & -4.018 & -2.223   \\
250 &  9.959 & -2.701 & -4.983 & -5.056 & -4.439   \\
    &        & -3.778 & -2.873 & -6.193 & -3.640   \\
\hline
\end{tabular}
}
\label{table:table33}
\end{center}
\end{table}

The most relevant part of the optical potential for the nucleus-nucleus elastic scattering is the tail of the potential, in particular at the strong 
absorption radius $R_{SA}$, which is closely related to the reaction cross-section ( see for instance \cite{bhagwat09} and references therein). In 
Table \ref{table:table33} we give the $R_{SA}$ values corresponding to the reactions and energies displayed in Figures \ref{fig1} and \ref{fig1a} 
obtained using the phenomenological optical potentials that describe the elastic scattering of $^2$H \cite{han06}, $^3$H and 
$^3$He \cite{pang09}and $^4$He \cite{su15}. 
The $R_{SA}$ is defined as the distance of the closest approach of the trajectory with angular momentum $L$ corresponding to transparency 
function $\vert S_L \vert=0.5$, which in the low energy regime, as the one considered in this work, can be approximated by the closest distance 
reached by the Coulomb orbit with  $\vert S_L \vert=0.5$.  Notice that the $R_{SA}$ extracted with the theoretical models used in this work, 
namely KD, MOPG, and RMOPG, provide very close values to the ones reported in this Table. For the sake of clarity, we will use the $R_{SA}$ obtained
 from the phenomenological optical potential in what follows. In the same Table, we also show the values of the real and imaginary 
parts of the central term of the optical 
potential at $R_{SA}$ computed with the aforementioned phenomenological optical potential  and with the different theoretical models 
discussed in this work. From 
this Table, we can see that the values of the real and imaginary parts at the $R_{SA}$ predicted by the theoretical models, in particular by the KD 
and RMOPG ones, agree quite well among them. This fact suggests that the scattering observables computed with these models should be quite similar 
\cite{roubos06}. However, the optical potentials at $R_{SA}$ computed with the theoretical models differ more from the predictions of the 
phenomenological ones. 
This implies that the theoretical predictions, obtained with global models,  will reproduce the experimental data less accurately than the local 
phenomenological optical potentials specifically designed for describing a given reaction. However, it should also be noted that the 
real and imaginary parts of the optical 
potentials for describing light particle scattering are quite deep and vary rapidly with the distance. Therefore the differences in Table 
\ref{table:table33} could have less impact on the calculation of scattering observables that appears at first sight as we will see in the next 
discussions.

\subsection{Angular distributions}

\begin{figure}[ht]
\centering
\includegraphics[width=1.\linewidth]{ 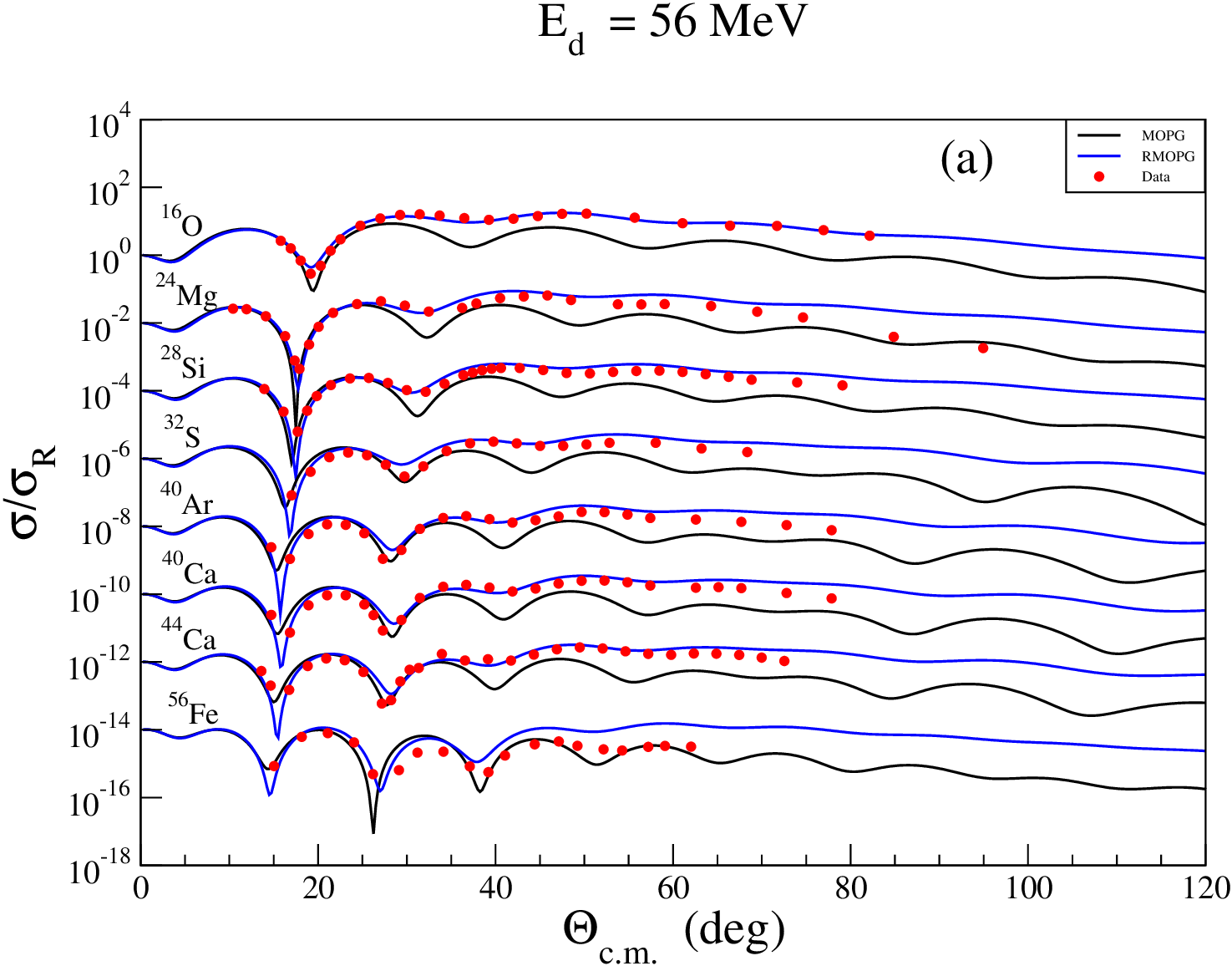} \\
\vspace{1cm}
\includegraphics[width=1.\linewidth]{ 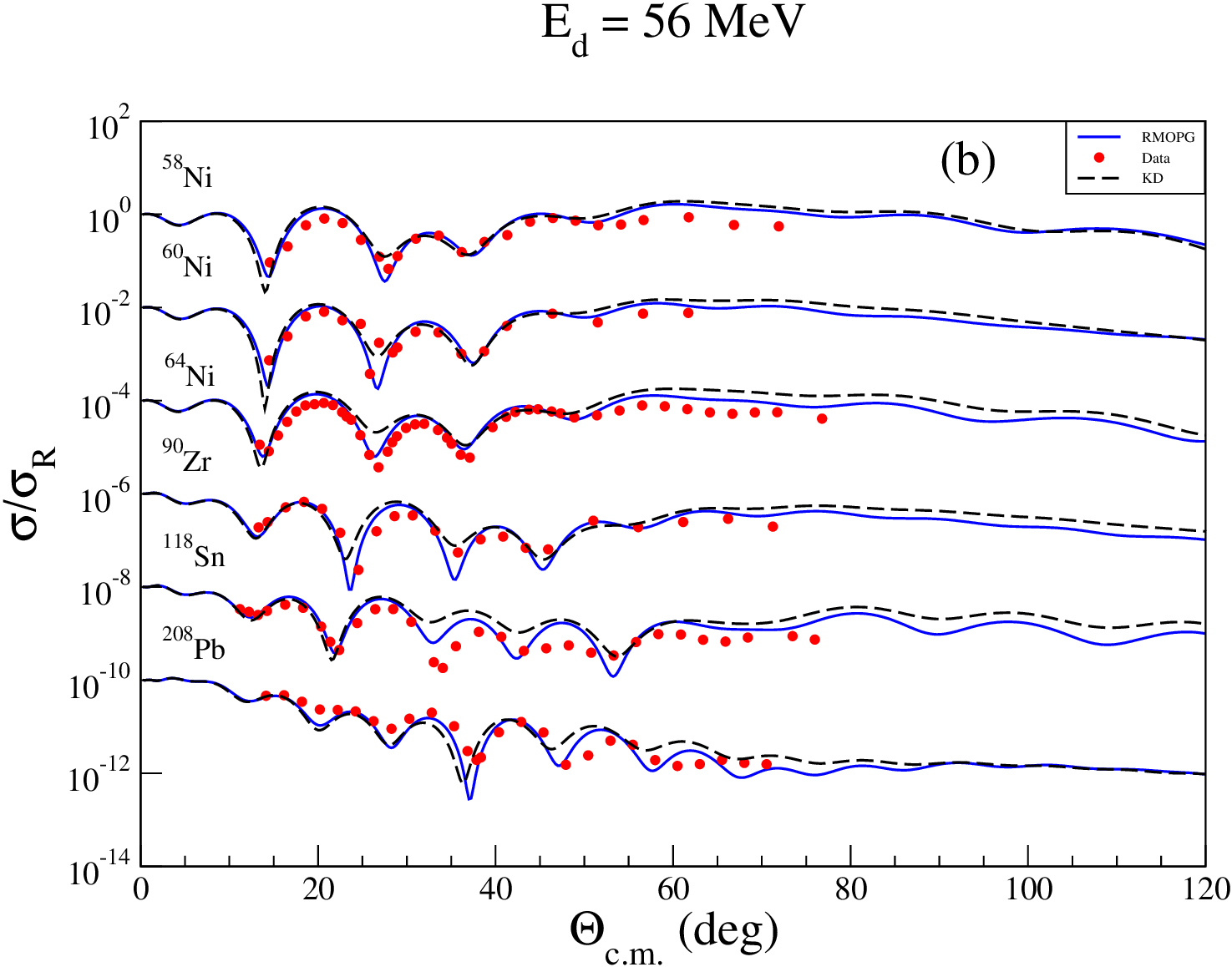}
\caption{Angular distributions in Rutherford unit of the elastic scattering of deuterons with an incident energy of 56 MeV by several target 
nuclei from $^{16}$O to $^{208}$Pb predicted by the RMOPG model compared to the MOPG (a) and KD (b) results. The experimental data are taken 
from \cite{hatanaka80}. Note that except for the topmost one, each angular distribution are offset by factors 100 from the preceding one.}
\label{fig3}
\end{figure}

We want now to investigate the predictive power of the light particle-nucleus microscopical optical potential based of 
the Gogny interaction derived in this work. To get some insight about the dependence of the angular distributions on the mass 
of the target  and the energy of the projectile,
we display in Figures \ref{fig3}, \ref{fig9} and \ref{fig12} the elastic scattering angular distributions 
in Rutherford units of incident 
$^2$H at 56 MeV, $^3$H at 33 MeV, $^3$He at 119 MeV and $^4$He at 35 and 104 MeV from different targets. Globally, we see from these Figures that 
the MOPG, which does not contain free parameters fitted to scattering data, reproduces the experimental results in a quite satisfactory way. 

In the case of elastic scattering of deuterons at 56 MeV on targets from $^{16}$O to $^{208}$Pb, we see that our renormalized model RMOPG reproduces
 very nicely the experimental behavior \cite{hatanaka80} of not only the dips at low scattering angles but also for larger scattering  
angles. We also see in the upper panel that the unrenormalized model MOPG is also in good agreement with the experiment in the dips but for larger
 scattering angles predicts a slightly smaller differential cross-section than the experimental data and shows a smooth oscillatory trend in 
disagreement with the experiment. In the lower panel, we compare the RMOPG and KD results. We see that both models show almost
an identical behavior reproducing fairly well the experimental data with few exceptions.
     
\cite{hatanaka80}
\begin{figure}[ht]
\centering
\includegraphics[width=1.0\linewidth]{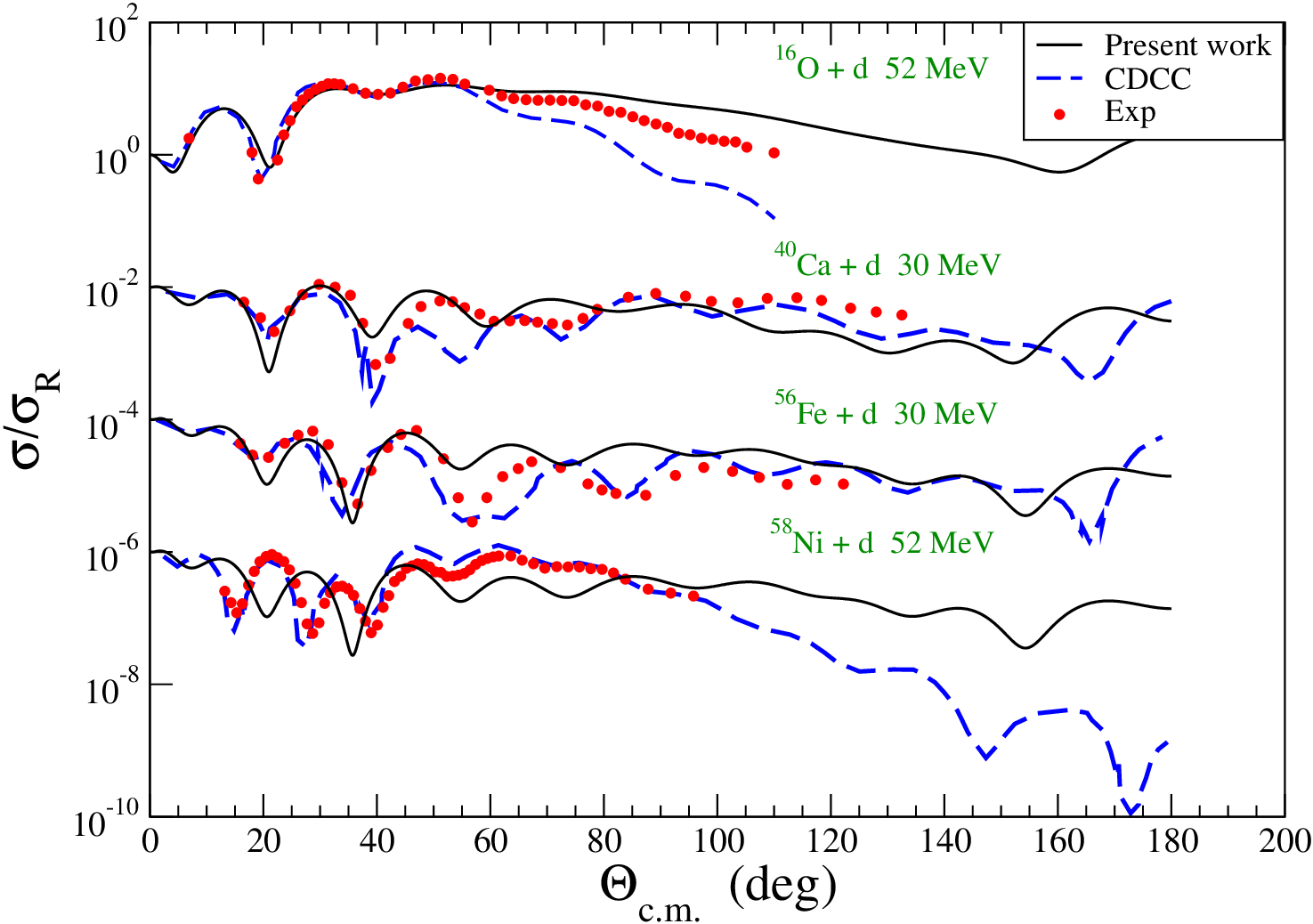}\\
\caption{Angular distributions in Rutherford units of the elastic scattering of deuterons on different targets obtained using the KD model. 
Solid and dashed lines are the results predicted by the approach used in the present work and by the CDCC of Ref.\cite{chau06}.
Experimental values are taken from the EXFOR Database \cite{exfor}. Note that except for the topmost one, each angular distribution 
are offset by factors 100 from the preceding one.}
\label{cdcc}
\end{figure}

As it is known, the breakup of loosely bound projectiles, such as $^2$H, $^3$H, or $^3$He,  has a relevant impact on the elastic 
scattering of such types of projectiles. For the reasons pointed out before, in the present work we have used the simplest approach proposed in the 
original paper of Watanabe, which includes the breakup effects as a whole in the imaginary part of the optical potential \cite{watanabe58}. However,
 in order to assess the reliability of our approach, we display in Figure \ref{cdcc} the differential cross sections obtained with our approach 
(solid line) and with the CDCC method (dashed line) using in both calculations the KD model. From this figure, we see that the predictions of our 
approach are reasonable, reproducing quite well the experimental data and with an overall agreement with the CDCC results taken 
from Ref.\cite{chau06}, which, as expected, reproduce slightly better the experimental values. 

\begin{figure}[ht]
\centering
\includegraphics[width=1.\linewidth]{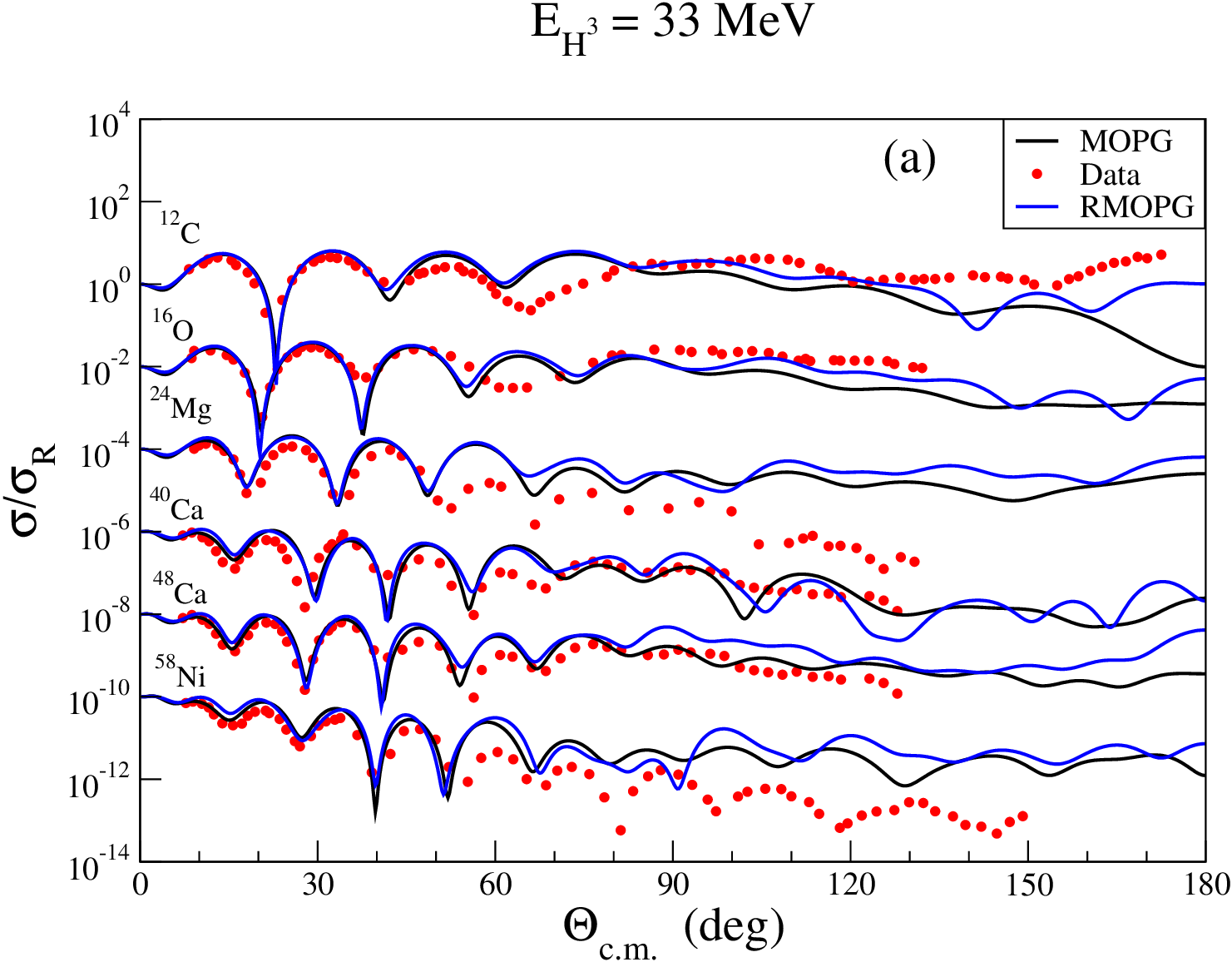}\\
\vspace{1cm}
\includegraphics[width=1.\linewidth]{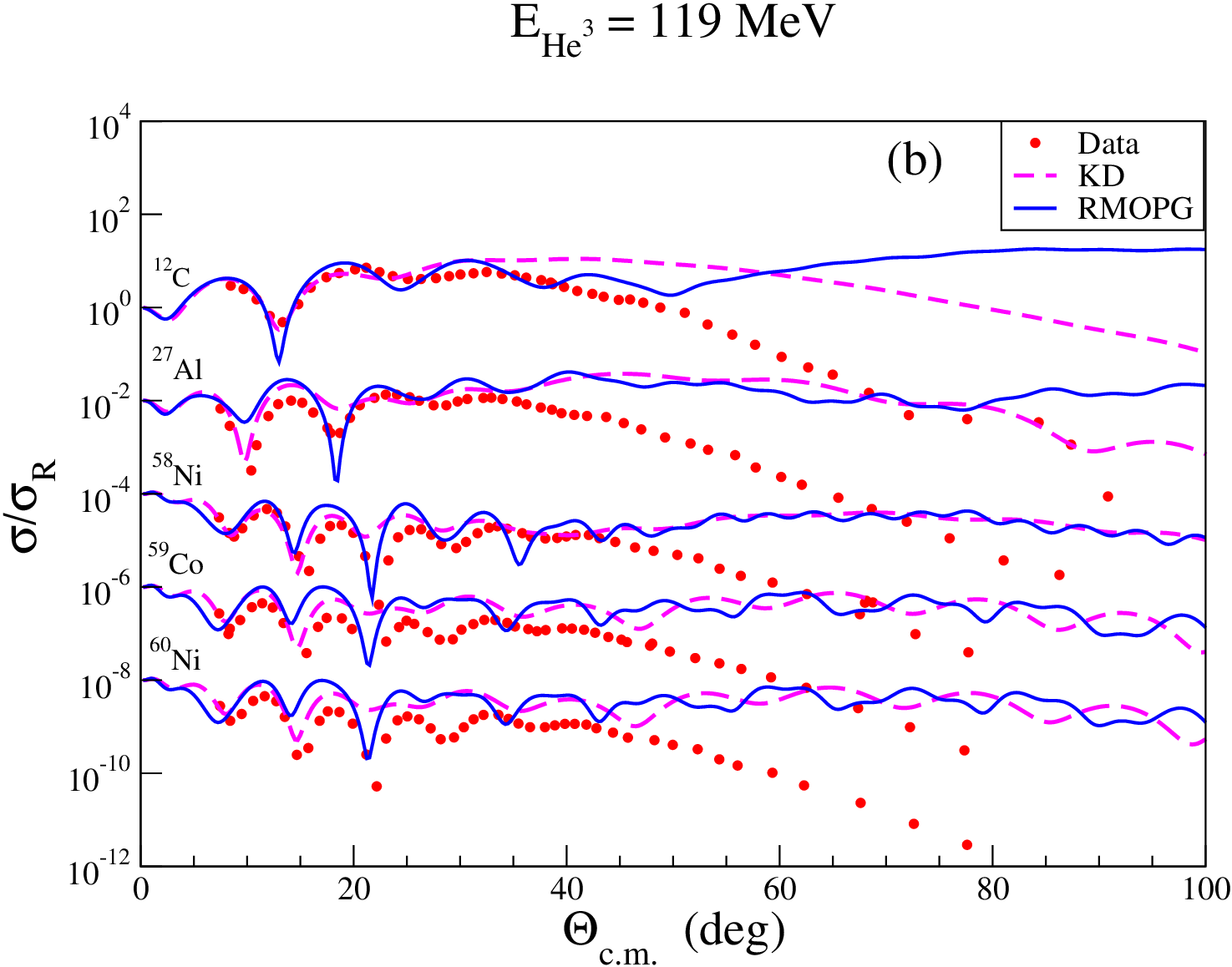}
\caption{Angular distributions in Rutherford units of the elastic scattering of $^3$H with incident energy of 33 MeV from several targets from 
$^{12}$C to $^{58}$Ni (a) and of  $^3$He with incident energy of 119 MeV from $^{12}$C to $^{60}$Ni (b). The experimental data 
are taken from \cite{england87} and \cite{hyakutake78} for $^3$H and $^3$He, respectively. For useful comparisons, we also display the upper 
(lower) panel the theoretical results computed  with the RMOPG and MOPG (RMOPG and KD) models. Note that except for the topmost one, 
each angular distribution is offset by factors 100 from the preceding one.}
\label{fig9}
\end{figure}

In the two panels of Fig.\ref{fig9}, we display the angular distributions of incident $^3$H and $^3$He at energies of 33 MeV and 119 MeV, 
respectively, scattered by several light and medium mass targets from C to Ni. For $^3$H scattering the MOPG and RMOPG models predict similar 
results for all the considered reactions \cite{england87}. Both models reproduce reasonably well the experimental dips up to a 
scattering angle of about 50$^o$. Beyond this angle, the experimental behavior is qualitatively reproduced in the case of the 
$^{12}$C, $^{16}$O, $^{40}$Ca and $^{48}$Ca targets and less in the case of $^{24}$Mg and $^{58}$Ni, where a lack of absorption 
in the two theoretical calculations is observed. The predictions of our model for the scattering of $^3$He projectiles also reproduce the 
experimental data \cite{hyakutake78} at low scattering angles in a reasonable way. Again the position of the dips in
 the angular distributions is predicted quite well by the MOPG and RMOPG models. Also, the experimental behavior is reasonably 
well averaged by our theoretical calculation up to scattering angles of 40$^o$-50$^o$. From these angles on, the experimental data 
exhibit a 
decreasing trend, which is not reproduced by our models. It is interesting to note that the behavior of the differential cross sections for 
scattering angles larger than $\approx$ 50$^o$ is not reproduced by the KD either, although it is predicted by the phenomenological potential that 
describes the elastic scattering of $A$=3 projectiles by heavier targets \cite{pang09}.

\begin{figure}[ht]
\centering
\includegraphics[width=1.\linewidth]{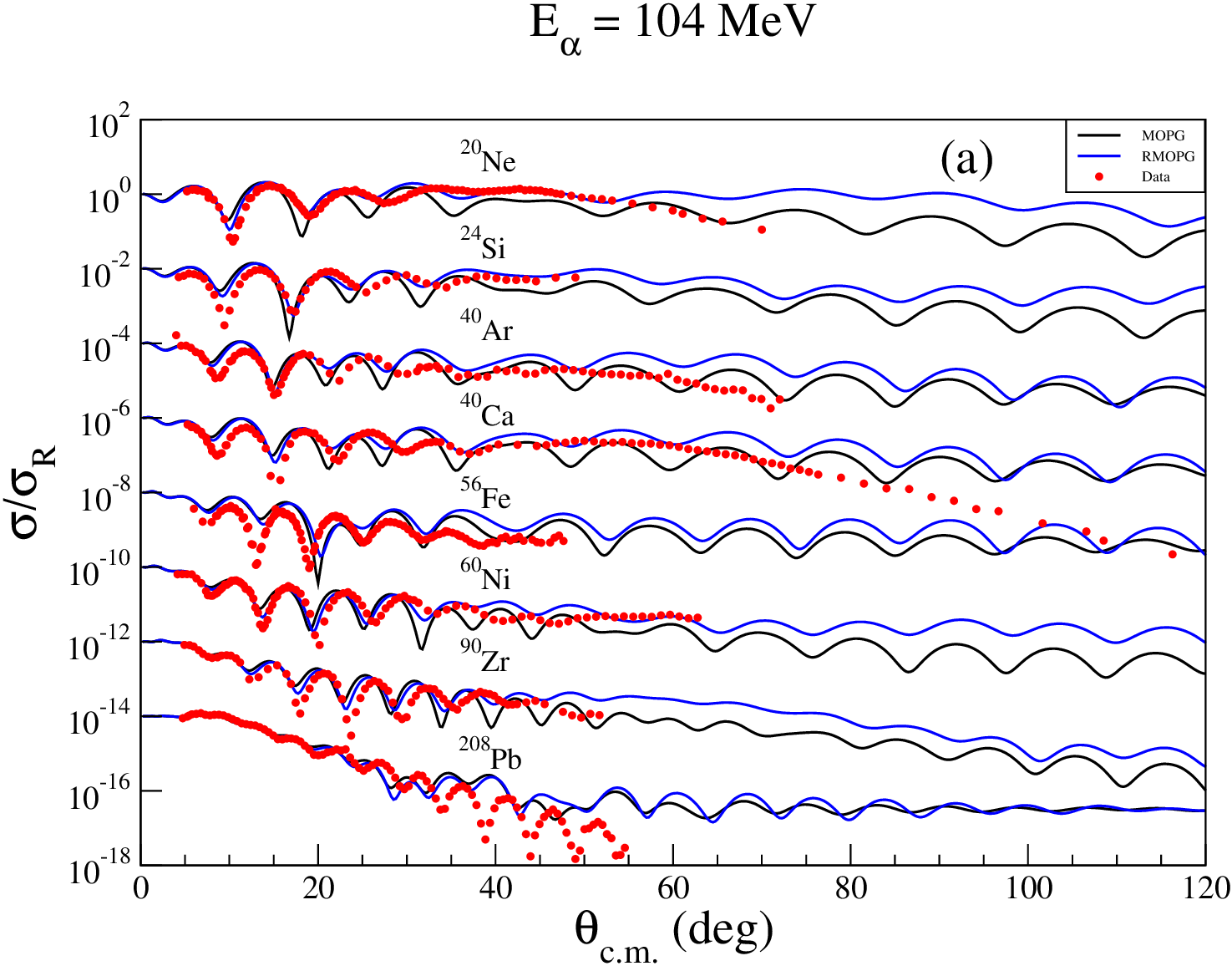}\\
\vspace{1cm}
\includegraphics[width=1.\linewidth]{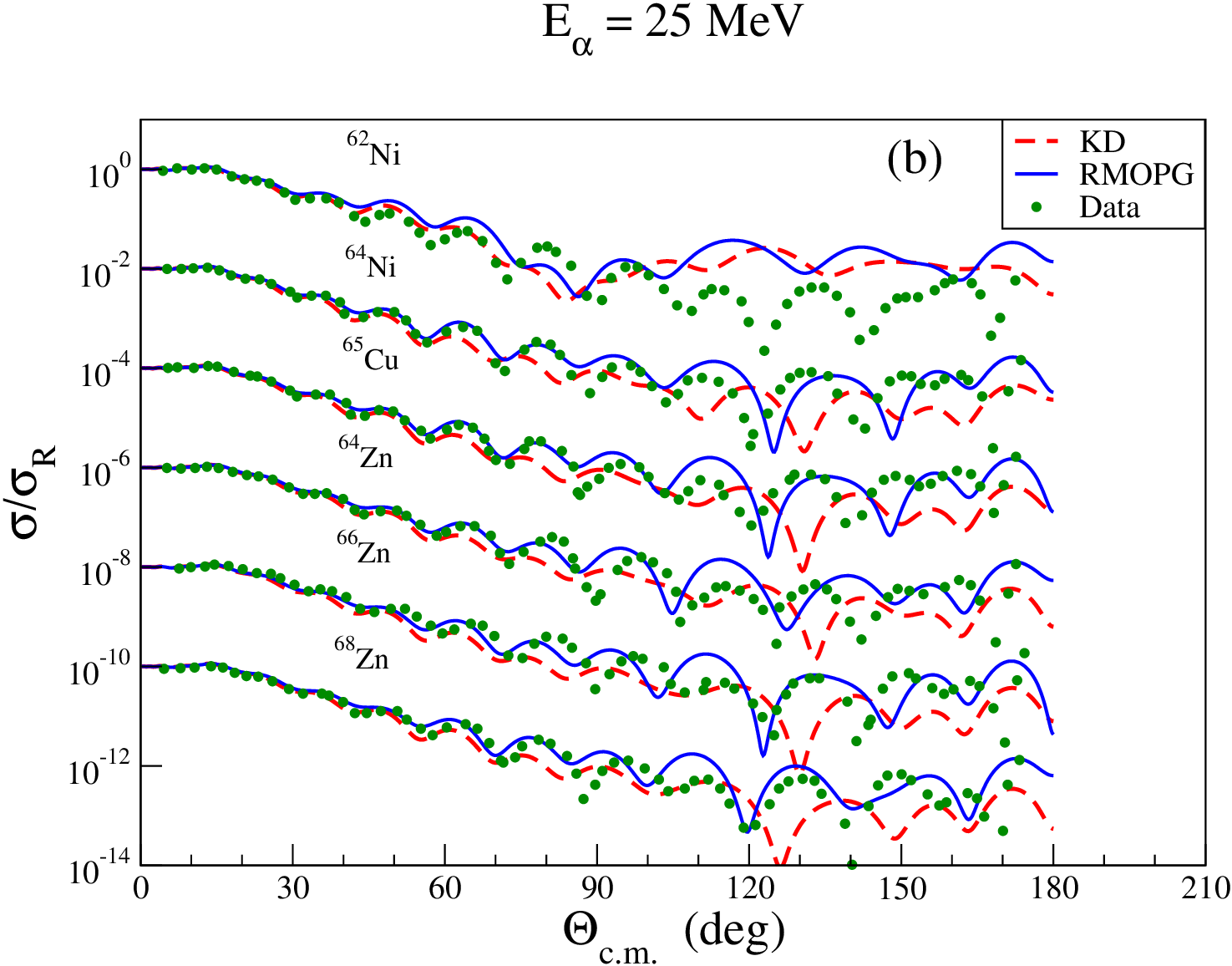}
\caption{(a): Angular distributions in Rutherford units of the elastic scattering of $\alpha$-particles  with incident energy of 104 MeV 
by several target nuclei from $^{40}$Ca to $^{208}$Pb predicted by the MOPG and RMOPG models in comparison with the experimental results 
\cite{hauser69,rebel72a,rebel72b,gils75}. (b): Angular distributions in Rutherford units of the elastic scattering of $\alpha$-particles 
of 25 MeV from several targets of mass number $A \approxeq$ 60. The experimental data are taken from \cite{england82}. 
Note that except for the topmost one, each angular distribution are offset by factors 100 from the preceding one.}
\label{fig12}
\end{figure}

In the upper panel of Figure \ref{fig12} we display the angular distributions in Rutherford units corresponding to the 
elastic scattering of $\alpha$-particles by different target 
nuclei at a given incident energy of 104 MeV \cite{hauser69,rebel72a,rebel72b,gils75} computed with the MOPG and RMOPG models. We see that these 
two theoretical calculations predict quite similar differential cross sections, which describe reasonably well the experimental data except for 
the $^{208}$Pb target, where the dips at scattering angles larger than 40$^o$ are not reproduced. In the lower panel of this Figure \ref{fig12} 
we display angular distributions relative to Rutherford scattering of $\alpha$-particle scattering on different targets of mass number $A \approx 60$ 
computed with the RMOPG and KD models. Both theoretical calculations predict very similar results, which reproduce rather well the 
experimental data up to scattering angles of about 90$^o$. From this angle on, the agreement between the theoretical predictions and the experiment 
slightly deteriorates.

\begin{figure}[ht]
\centering
\includegraphics[width=1.\linewidth]{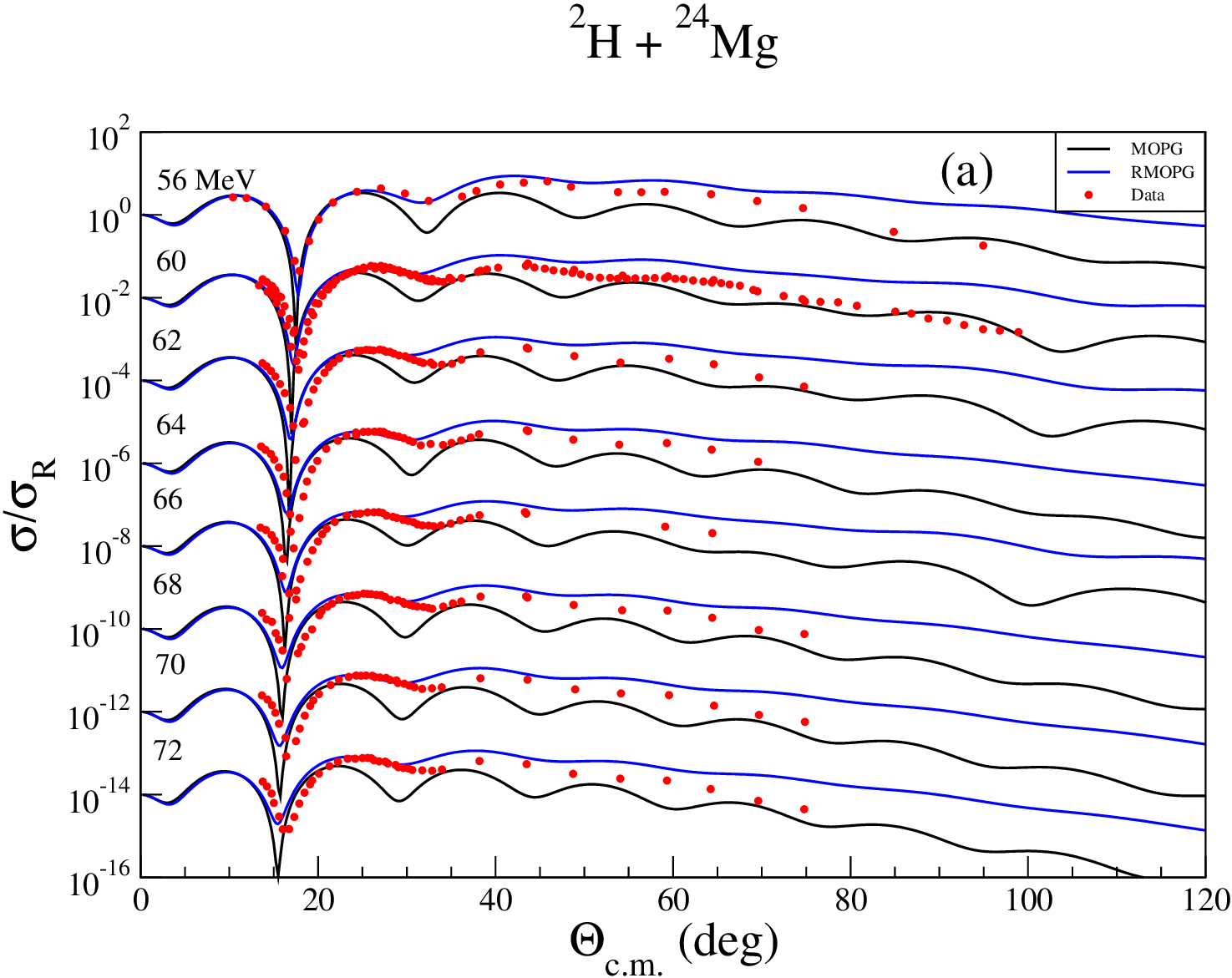}\\
\vspace{1cm}
\includegraphics[width=1.\linewidth]{ 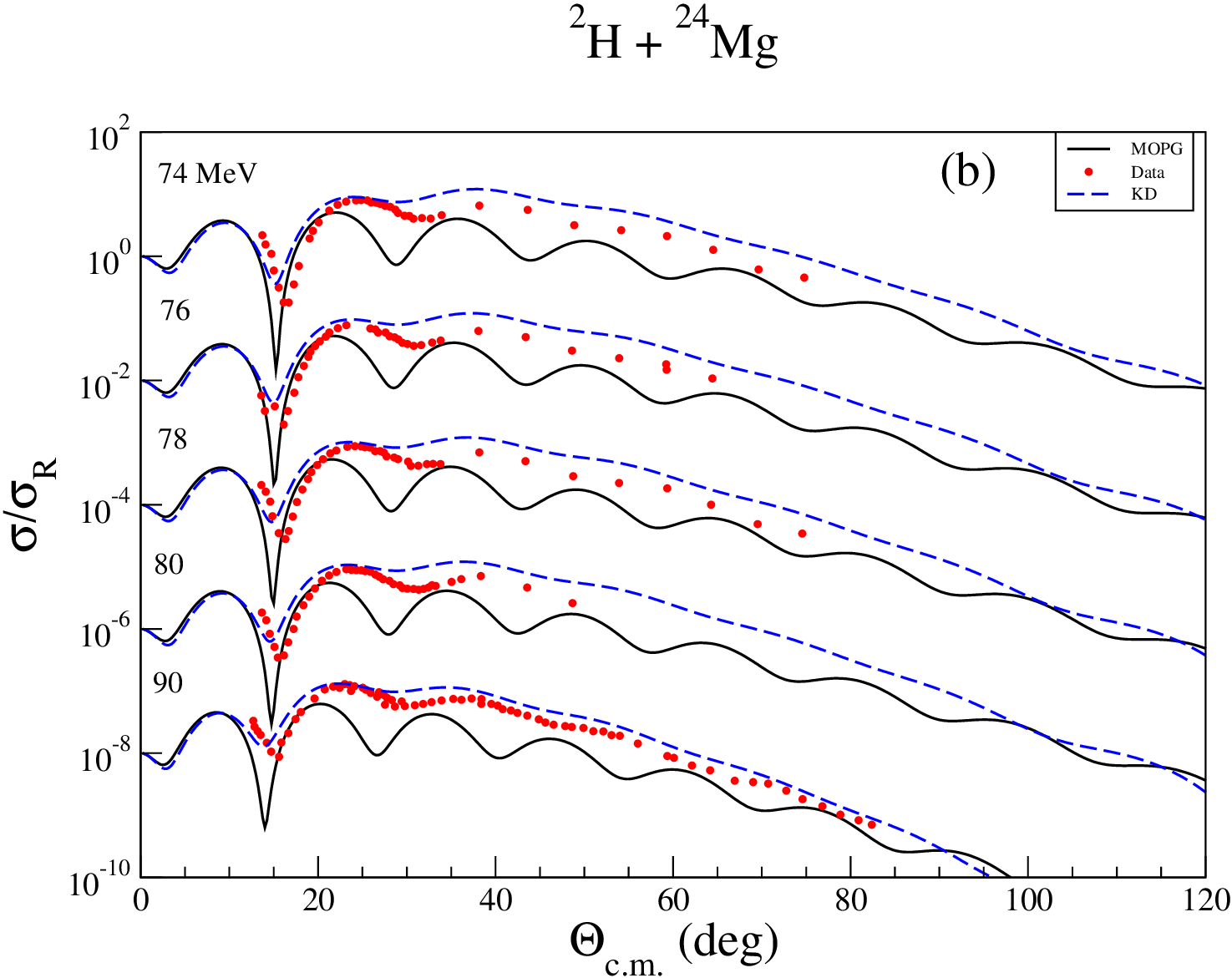}
\caption{Angular distributions in Rutherford units of the elastic scattering of deuterons from $^{24}$Mg at different incident energies of the 
projectile computed with the MOPG and RMOPG models (a) and with the MOPG and KD models (b) compared to the experimental values 
\cite{kiss76,baumer01}. Note that except for the topmost one, each angular distribution are offset by factors 100 from the preceding one.}
\label{fig4}
\end{figure}

Next, we want to investigate the energy dependence of the differential cross sections for a given target. To this end, we display in Figure 
\ref{fig4} the angular distributions of deuterons scattered by a target of $^{24}$Mg in the range of energies between 56 and 72 MeV in the upper 
panel and between 74 and 90 MeV in the lower panel. In the case of elastic scattering of deuterons in the range between 56 and 72 MeV, the MOPG 
and RMOPG predict quite well the position and depth of the diffraction minima. For larger scattering angles the experimental data decrease. 
This trend is well reproduced by the RMOPG model while MOPG shows a wiggly pattern with a value slightly smaller than the 
experimental data due to the large absorption of this model. In the energy range between 74 and 90 MeV, which is  displayed in the 
lower panel, we see that the KD model reproduces nicely the experimental data with a quality similar to that of the RMOPG in the upper panel. 
The MOPG predictions, which show the same trends as in the upper panel, reproduce the decreasing behavior of the experimental values 
only in a qualitative way.

\begin{figure}[ht]
\centering
\includegraphics[width=0.9\linewidth]{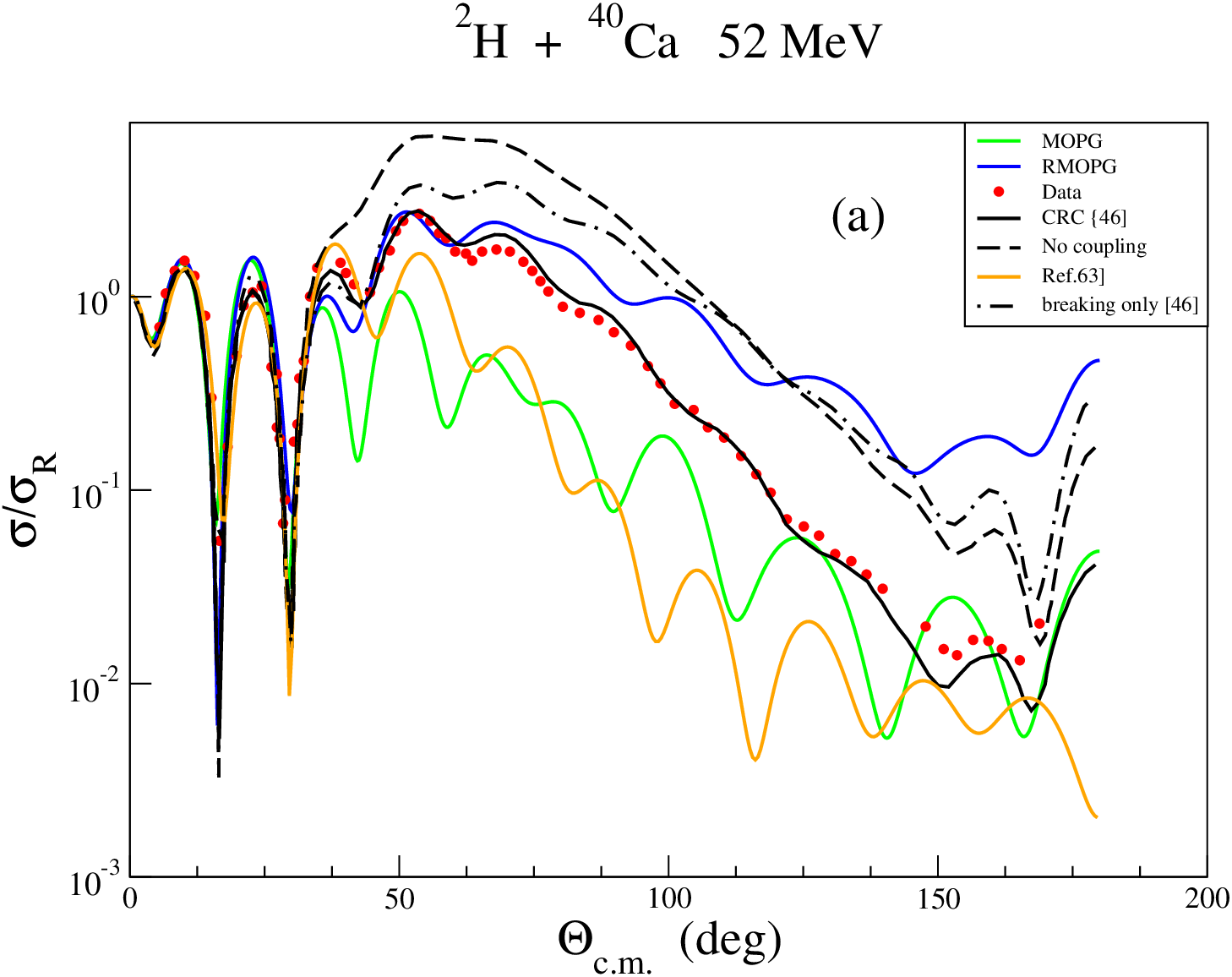}\\
\vspace{1cm}
\includegraphics[width=0.9\linewidth]{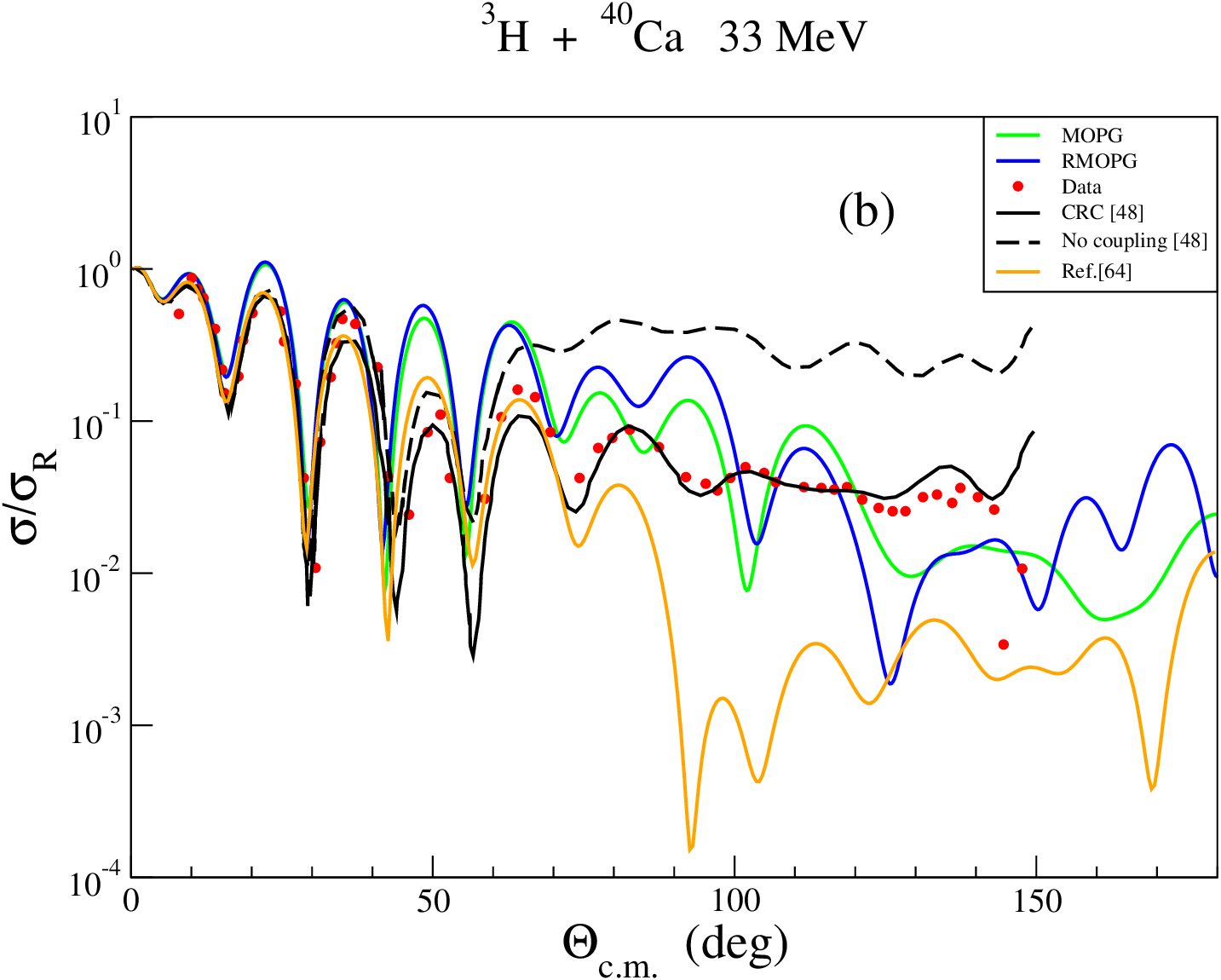}
\caption{(a): Angular distributions in Rutherford units of the elastic scattering of deuteron  with an incident energy of 52 MeV
by a target of $^{40}$Ca predicted by the MOPG and RMOPG models in comparison with the results of the CDCC and CRC calculations of
Keeley and Mackintosh reported in \cite{keeley08}. The results provided by the phenomenological potential of Ref.~\cite{han06} are also displayed. 
(b): The same as in the upper panel for the scattering of triton at 33 MeV by  a $^{40}$Ca target. The CDCC and CRC are taken from 
\cite{keeley20} and the phenomenologoical predictions from Ref.~\cite{pang09}}.
\label{fig13}
\end{figure}

Although our model is able to describe deuteron, triton, helion and $\alpha$-particle scattering by heavier nuclei for scattering
angles up to $\approx$ 40$^o$-50$^o$ in a rather reasonable way, it fails for larger angles where discrepancies with the experimental values can be 
important. Overall our model is not completely satisfactory due to the following reasons. 
On the one hand, our global optical potentials for describing light particle scattering, namely MOPG and 
RMOPG,  are built up with the Watanabe model using the underlying microscopic nucleon-nucleus optical potential derived in Ref.\cite{lopez21}. 
The former does not contain parameters adjusted to scattering data and the latter has been renormalized using experimenral nucleon-nucleus 
scattering results. Therefore neither of the two models has been fitted to experimental data of light particle-nucleus scattering and consequently, 
our models are fully predictive in this respect. Thus it is not surprising that the predictive power of the MOPG and RMOPG be less than the one of 
phenomenological optical models because in the latter case their parameters are fitted to reproduce the scattering data of a given projectile. 
On the other hand, there are the intrinsic limitations of the folding model, which does not take into account the well-established coupled channel 
and coupled reaction channels that are very important
for an accurate description of the light particle scattering \cite{keeley08,mackintosh19,keeley20,keeley23}. Calculations including coupling 
collective states and pickup reactions allow to extract  the so-called dynamical polarization potential by inversion of the elastic channel 
$S$ matrix. This potential added to the folding potential provides a very accurate description of the angular distribution of the elastic 
cross-section \cite{keeley08,mackintosh19,keeley20,keeley23}. In Figure \ref{fig13} we display the angular distributions relative to Rutherford of 
the $^2$H + $^{40}$Ca at 52 MeV and $^3$H + $^{40}$Ca at 33 MeV reactions computed with the coupled reaction channel method in Refs.~\cite{keeley08} 
and \cite{keeley20}, respectively, together with the predictions of our MOPG and RMOPG models. We see that all the folding models displayed in these 
figure, including the ones of Refs.~\cite{keeley08} and \cite{keeley20}, describe the experimental data up to scattering angles about 40$^o$-50$^o$
 and predict very different values for larger angles, which may indicate a failure of the pure folding models. However, when the dynamical 
polarization potential is added to the folding contribution, the experimental angular distributions are reproduced with very high precision as can 
be seen in Figure \ref{fig13} and in the results reported in \cite{keeley08,mackintosh19,keeley20,keeley23} and references therein. It is important
 to point out that the effects of the dynamical polarization potential cannot be recovered by renormalization of the folding potential as discussed
 in \cite{keeley23}. In Figure \ref{fig13} we have also plotted the angular distributions for the above-mentioned reactions computed with the 
phenomenological optical potentials for deuterons \cite{han06} and for tritons \cite{pang09}. We see that these phenomenological optical potentials
 describe again the  experimental data up to 40$^o$-50$^o$ and fail forlarger angles showing that this type of global potentials are not very well 
suited to deal with scattering on closed-shell targets \cite{pang09,mackintosh19}.

\section{Reaction cross sections}

\begin{table}[ht]
\begin{center}
\caption{Reaction cross sections for $^2$H-nucleus scattering computed with the MOPG, RMOP and KD models computed with the approach used in this 
work. Some results from the more elaborated CDCC method reported in \cite{chau06} are also given. Experimental values 
are taken from Ref.~\cite{auce96}}
\scalebox{0.9}{
\begin{tabular}{cccccc}
\cline{1-6}
E & Data(mb) & MOPG(mb) & RMOPG(mb) & KD(mb) & CDCC \cite{chau06} \\
\hline
\multicolumn{5}{c}{$^{16}$O}\\
\hline
37.9 & 962$\pm$27 & 1136 & 1074 & 1066 & 1122 \\
65.5 & 811$\pm$19 & 1067 &  907 &  883 &  943 \\
97.4 & 726$\pm$21 & 1011 &  822 &  745 &  790 \\
\hline
\multicolumn{5}{c}{$^{40}$Ca}\\
\hline
37.9 & 1439$\pm$43 & 1568 & 1503 & 1487 & 1590 \\
65.5 & 1338$\pm$28 & 1556 & 1391 & 1361 & 1424 \\
97.4 & 1260$\pm$30 & 1486 & 1274 & 1175 & 1244 \\
\hline
\multicolumn{5}{c}{$^{58}$Ni}\\
\hline
37.9 & 1625$\pm$51 & 1741 & 1678 & 1731 & 1824 \\
65.5 & 1571$\pm$33 & 1724 & 1559 & 1574 & 1683 \\
97.4 & 1524$\pm$45 & 1693 & 1479 & 1424 & 1503 \\
\hline
\multicolumn{5}{c}{$^{120}$Sn}\\
\hline
37.9 & 2240$\pm$69 & 2317 & 2243 & 2247 & 2341 \\
65.5 & 2346$\pm$51 & 2453 & 2286 & 2242 & 2330 \\
97.4 & 2351$\pm$55 & 2426 & 2193 & 2070 & 2176 \\
\hline
\multicolumn{5}{c}{$^{208}$Pb}\\
\hline
37.9 & 2844$\pm$142 & 2643 & 2576 & 2676 & 2736 \\
65.5 & 3049$\pm$71 & 2976 & 2809 & 2865 & 2937 \\
97.4 & 3250$\pm$82 & 3043 & 2797 & 2781 & 2881 \\
\hline
\end{tabular}}
\label{table:table12}
\end{center}
\end{table}
\begin{figure}[ht]
\centering
\includegraphics[width=1.\linewidth]{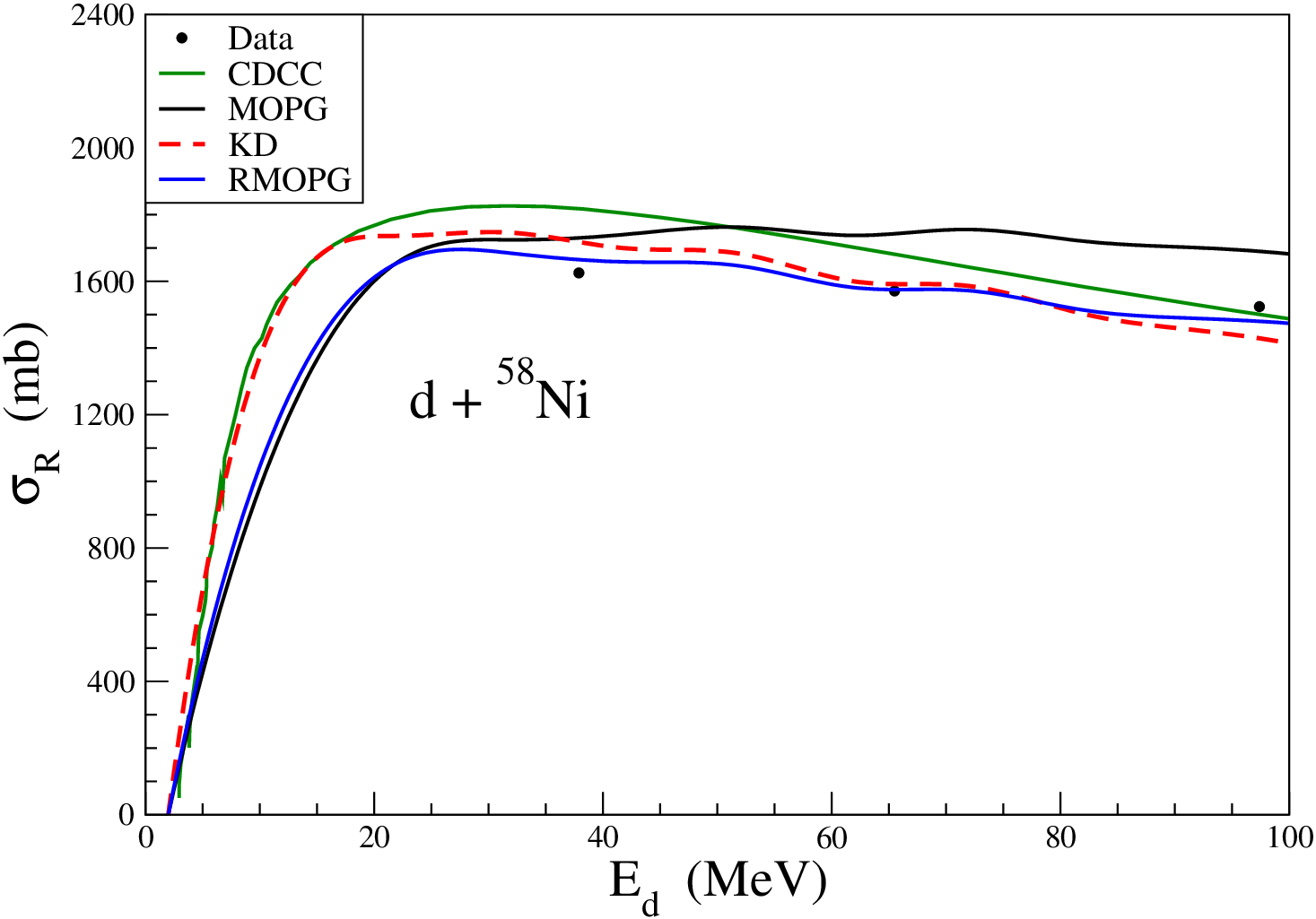}\\
\vspace{1cm}
\includegraphics[width=1.\linewidth]{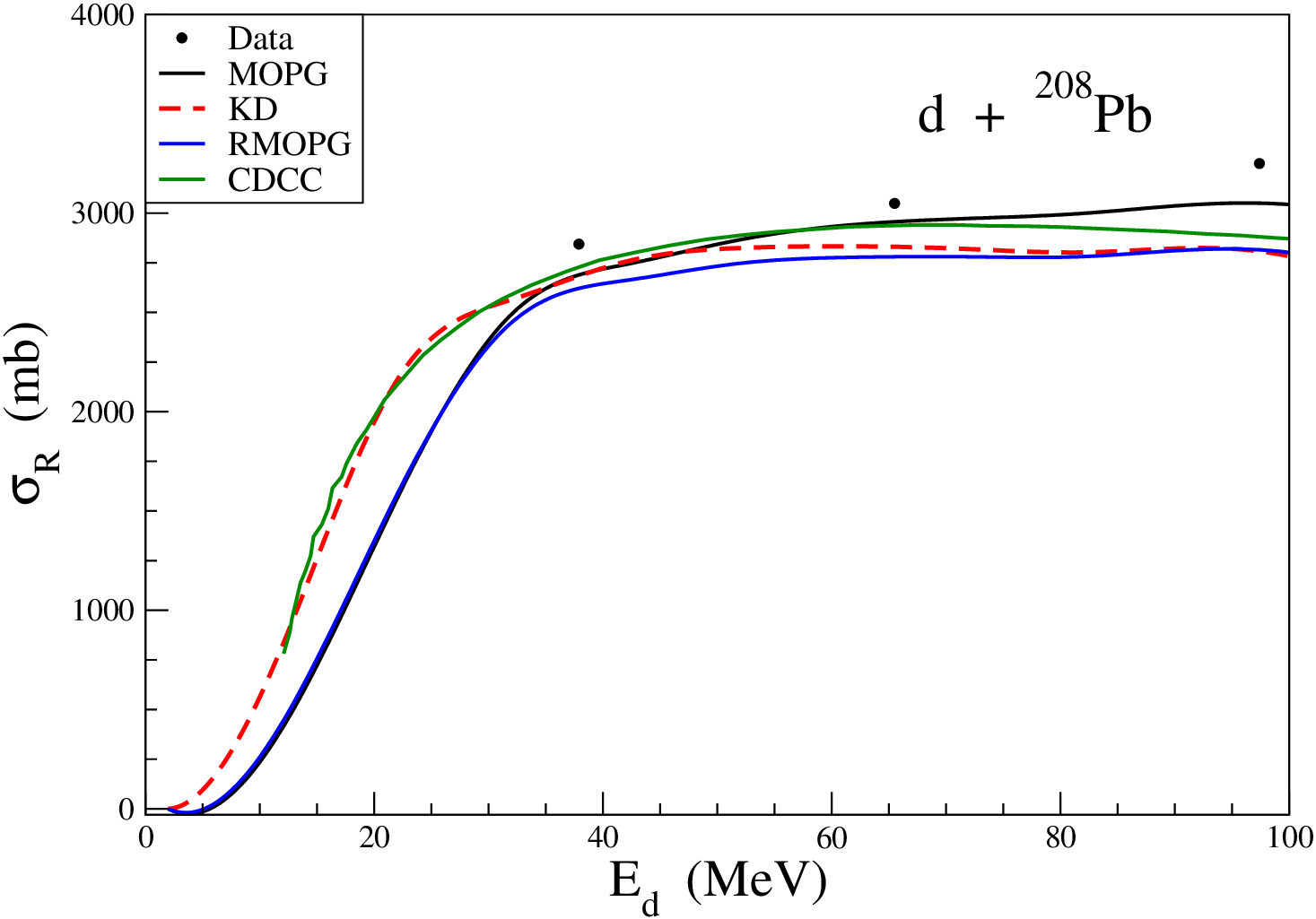}
\caption{Reaction cross sections as a function of the energy of the incident deuteron on $^{58}$Ni (up) and $^{208}$Pb (down) computed with the 
approach developed in this work using the MOPG, RMOPG and KD models. The predictions of the CDCC method taken from \cite{chau06} are also displayed.}
\label{fig20}
\end{figure}
\begin{table}[ht]
\begin{center}
\caption{Reaction cross sections for $^3$He-nucleus scattering computed with the MOPG, RMOPG,  and KD models. Experimental values are taken from 
Ref.~\cite{ingemarsson01}}
\begin{tabular}{ccccc}
\cline{1-5}
E & Data(mb) & MOPG(mb) & KD(mb) & RMOPG(mb) \\
\hline
\hline
\multicolumn{5}{c}{$^{16}$O}\\
\hline
96.4 & 975$\pm$35 & 1163 & 1045 & 1044 \\
137.8 & 850$\pm$50 & 1115 & 908 & 960 \\
167.3 & 800$\pm$25 & 1090 & 844 & 939 \\
\hline
\multicolumn{5}{c}{$^{40}$Ca}\\
\hline
96.4 & 1360$\pm$90 & 1599 & 1459 & 1462 \\
137.8 & 1280$\pm$85 & 1557 & 1333 & 1386 \\
167.3 & 1225$\pm$75 & 1524 & 1256 & 1353 \\
\hline
\multicolumn{5}{c}{$^{58}$Ni}\\
\hline
96.4 & 1690$\pm$100 & 1771 & 1681 & 1635 \\
137.8 & 1570$\pm$80 & 1806 & 1646 & 1542 \\
167.3 & 1470$\pm$75 & 1734 & 1500 & 1533 \\
\hline
\multicolumn{5}{c}{$^{120}$Sn}\\
\hline
96.4 & 2285$\pm$165 & 2458 & 2263 & 2305 \\
137.8 & 2230$\pm$100 & 2552 & 2235 & 2316 \\
167.3 & 2180$\pm$100 & 2523 & 2158 & 2245 \\
\hline
\multicolumn{5}{c}{$^{208}$Pb}\\
\hline
96.4 & 2765$\pm$250 & 2913 & 2823 & 2757 \\
137.8 & 2850$\pm$250 & 3081 & 2876 & 2822 \\
167.3 & 2820$\pm$180 & 3124 & 2891 & 2680 \\
\hline
\end{tabular}
\label{table:table13}
\end{center}
\end{table}

\begin{table}[ht]
\begin{center}
\caption{Reaction cross sections for $^4$He-nucleus scattering computed with the MOPG, RMOPG, and KD models. Experimental values are taken from 
Ref.~\cite{ingemarsson00}}
\scalebox{0.9}{
\begin{tabular}{ccccc}
\cline{1-5}
E & Data(mb) & MOPG(mb) & KD(mb) & RMOPG(mb) \\
\hline
\multicolumn{5}{c}{$^{16}$O}\\
\hline
117.2 & 973$\pm$62 & 1162 & 1076 & 1058 \\
163.9 & 895$\pm$100 & 1134 & 990 & 986 \\
192.4 & 850$\pm$58 & 1118 & 945 & 955 \\
\hline
\multicolumn{5}{c}{$^{40}$Ca}\\
\hline
117.2 & 1470$\pm$60 & 1614 & 1512 & 1494 \\
163.9 & 1410$\pm$120 & 1579 & 1463 & 1360 \\
192.4 & 1370$\pm$70 & 1599 & 1393 & 1389 \\
\hline
\multicolumn{5}{c}{$^{60}$Ni}\\
\hline
117.2 & 1670$\pm$85 & 1832 & 1774 & 1711 \\
163.9 & 1700$\pm$160 & 1850 & 1713 & 1659 \\
192.4 & 1610$\pm$90 & 1828 & 1649 & 1609 \\
\hline
\multicolumn{5}{c}{$^{120}$Sn}\\
\hline
117.2 & 2360$\pm$150 & 2472 & 2342 & 2334 \\
163.9 & 2380$\pm$250 & 2516 & 2285 & 2308 \\
192.4 & 2300$\pm$170 & 2550 & 2276 & 2267 \\
\hline
\multicolumn{5}{c}{$^{208}$Pb}\\
\hline
117.2 & 2990$\pm$180 & 2972 & 2940 & 2829 \\
163.9 & 2720$\pm$250 & 2987 & 2915 & 2664 \\
192.4 & 3900$\pm$190 & 2989 & 2871 & 2507 \\
\hline
\end{tabular}}
\label{table:table14}
\end{center}
\end{table}

The reaction cross-section measures the flux of incident particles that are removed from the elastic channel because of the non-elastic process. 
In addition to  some technological applications, the main interest of the study of reaction cross sections lies in the fact that they can be very 
useful in the analysis of angular distributions in elastic scattering in order to eliminate ambiguities of the optical potential 
\cite{ingemarsson01}. The study of non-elastic reactions requires accurate optical potentials because the imaginary part of their phase shifts 
determines not only the reaction cross sections but also the amplitude of the partial waves entering in the distorted-wave calculations needed to 
describe the inelastic process. Also, the reaction cross sections may be an important quantity to give global insight into analyzing the predictive 
power of different optical models. In this section we compare the theoretical reaction cross sections obtained using the MOPG, RMOPG, and KD models
 for the scattering of $^2$H at 38, 65, and 97 MeV; $^3$He at 98, 138, and 167 MeV and $^4$He at 117.2, 163.9 and 192.4 MeV from a set of targets 
ranging from $^{16}$O to $^{208}$Pb with the corresponding experimental values, which are taken from Refs.\cite{auce96}, \cite{ingemarsson01} and 
\cite{ingemarsson00}, respectively.
 
Our theoretical results are collected in Tables \ref{table:table12} for $^2$H, \ref{table:table13} for $^3$He and \ref{table:table14} for $^4$He. 
The experimental values of the reaction cross sections increase when the mass of the target grows. As a function of the energy, the lightest targets
 show, for all the projectiles, a decreasing tendency with increasing energy. These global trends with the mass of the target and the energy of the 
projectile are fulfilled rather well by our theoretical calculations. In a more quantitative way, we see that for all the projectiles the MOPG 
without renormalization overestimates the experimental values, except for reactions of $^2$H and $^4$He on $^{208}$Pb targets, as a consequence of 
the large imaginary part in the nucleon-nucleus optical potential \cite{lopez21}. The RMOPG and KD models predict cross sections that are quite 
similar between them and  slightly overcome the experimental values for the lightest targets, in particular for the smallest energies. For medium 
mass nuclei up to $^{120}$Sn, both models describe quite accurately the experimental reaction cross sections while for the heaviest target 
$^{208}$Pb both models underestimate the experimental values by an amount that can be about 25\% in the case of scattering of $^4$He. In the 
particular case of scattering of $^2$H,  we also collect in Table \ref{table:table12} the CDCC reaction cross sections reported in Ref.\cite{chau06}, 
which are also computed using the KD model. The CDCC results exceed the values calculated in this work by about 5\%, pointing out that our approach
 takes into account a large amount of the breakup effects, at least in what concerns reaction cross sections. In Figure \ref{fig20} we display the 
reaction cross section of the  $^2$H + $^{58}$Ni and  $^2$H + $^{208}$Pb reactions as a function of the energy of the projectile computed with all 
the models used in this work. In the case of $^{58}$Ni target, we see that the reaction cross section increases with the energy and reaches a maximum at
 about $E_d \approxeq$ 20 MeV and then decreases. However, for the $^{208}$Pb, the reaction cross section is an increasing function of the energy, at 
least in the considered range. We also can see that the CDCC predictions and our results computed in this work almost coincide up to an energy of 
the projectile $E_p$ about 20 MeV for the $^{58}$Ni and about 40 MeV for the $^{208}$Pb target. From these energies on, the CDCC prediction,
 is larger by $\approx$ 5\% than the result computed in this work as a consequence of the better description of the breakup effects. For both reactions
 and energies below 40-50 MeV, the KD cross sections are larger than the values predicted by the MOPG and RMOPG models and the opposite trend 
happens for higher energies. In this region and for the $^{58}$Ni target, the MOPG and CDCC results are larger than the experimental data, while 
the RMOPG and KD results, computed with the extended Watanabe approach used in this work, reproduce quite accurately the experimental values. 
For the $^{208}$Pb target the experimental reaction cross sections are underestimated by all the models considered, the MOPG without renormalization
 being the model that better agrees with the experimental data.

\section{Summary and conclusions}  

We have derived a microscopic optical potential to describe the elastic scattering of  light particles, such as deuterons, tritons, helions and 
$\alpha$-particles from heavier target nuclei. This optical potential is obtained through an extended  Watanabe model, whose basic 
ingredients are the neutron-nucleus and proton-nucleus optical potentials and the projectile wave functions. This simple model neglects the 
interaction among the nucleons of the projectile and assumes that effects from dissociation and distortion of the wave function of the projectile 
can be included phenomenologically in the imaginary part of the optical potential. Effects due to inelastic and pickup reactions 
have not been considered either in our simple model. The nucleon-nucleus optical potential used here was derived in 
a previous work within a semi-microscopic nuclear matter approach where the real and imaginary parts are given, respectively, by the first and 
second-order terms of the mass  operator, which is determined by means of a Brueckner-Hartree-Fock calculation using a $G$-matrix built up with an 
effective Gogny interaction. This nucleon-nucleus potential is supplemented by the Coulomb potential for incident protons and by a real spin-orbit 
potential obtained from a self-consistent quasi-local HF calculation in the target nucleus.

As is expected, for all the projectiles considered, the real and imaginary parts of the theoretical light particle-nucleus optical potential 
derived in this work decrease and increase, respectively, when the energy of the projectile grows. At low incident energy, the imaginary part is 
strongly peaked at the surface. When the energy increases the volume absorption grows and the relative surface absorption diminishes. 
The angular distribution of the elastic scattering of light particles at different energies from different target nuclei computed with the optical 
potential derived in this work exhibits reasonably agreement with the experimental data, although the quality depends on the reaction. In 
general, for scattering angles below 50$^o$, the diffraction pattern of the experimental data at high enough incident energy is quite well 
reproduced by our calculation as well as the position of the first dips. For larger scattering angles there are discrepancies between the values of
 the differential cross sections predicted by our model and the experimental values, pointing out to a too strong absorption for light targets 
although this trend is reversed for heavier nuclei.  The cross sections computed with our model decrease with the growing energy of the projectile 
for light and medium mass targets while the contrary happens for the heaviest target analyzed in this work, namely $^{208}$Pb, which is in 
agreement with the experimental trend. However, the values calculated with our MOPG model for all the targets except $^{208}$Pb overestimate the 
experimental values, especially in the case of the lightest targets with a mass numbers smaller than $A=40$.    

To have a better insight into the quality of the results obtained with the MOPG model, we have repeated the theoretical calculations within extended
Watanabe model but using the K\"oning-Delaroche nucleon-nucleus optical potential instead of the optical potential based on the Gogny interaction. 
These KD optical potentials for describing the elastic scattering of light particles predict angular distributions and reaction cross sections in 
better agreement with the experiment than our MOPG model. We have also used this KD model in the case of deuteron scattering to compare our extended 
Watanabe prescription with the results provided by the more elaborated CDCC method, which takes into account explicitly the breakup effects. We 
find that the CDCC method improves slightly the angular distributions and provides reaction cross sections about 6-7\% larger.    

To obtain a better description of light particle elastic scattering with our model, we improve the starting nucleon-nucleus potential 
by renormalizing with energy-dependent factors the real and imaginary parts of the central contribution and the spin-orbit potential following the 
protocol described in the Appendix. This renormalized  nucleon-nucleus potential also based on the Gogny interaction predicts a description of the 
nucleon-nucleus elastic scattering quite similar to the one provided by the K\"oning-Delaroche model in spite of the fact that the fitting procedure 
of both models is clearly different. The use of these renormalized nucleon-nucleus potentials in the extended Watanabe approach produces the RMOPG
model, which describes light  particle elastic scattering with a quality similar to that obtained with the KD model. In particular, the angular 
distributions computed with the RMOPG model reproduces the dips at small scattering angles and the exponential fall-off at large scattering data 
in much better agreement with the  experiment than the unrenormalized MOPG results. The reaction cross sections computed with the RMOPG are 
smaller than thones calculated with the unrenormalized MOPG model and quite similar to the values predicted by the KD model. 
We have also checked that the results discussed in this work are compatible with other folding
calculations using nucleon-nucleus microscopic optical models  built up with the Skyrme interaction. 

Overall our Watanabe model based on the Gogny force can provide a qualitative description of the light projectile-nucleus elastic scattering.
In particular, the angular distributions up to scattering angles about 40$^o$-50$^o$ reproduce the experimental data fairly well but fail   
for larger angles. Apart from its global charcter, the lack of accuracy in our model, as well as in any other obtained using  the folding method, is 
due to the absence of inelastic and reaction chanels in the elastic scattering calculation. A more accurate description of the systematic of the 
elastic scattering of light particle by heavier nuclei require local information about the structure of the target and nearby
nuclei, which is the necessary input for dealing with the more sophisticated CDCC and CRC calculations.
   
Although numerical results obtained with the 
Gogny models are available from the authors on request, an important and urgent task is to publish the numerical code. Work in this direction is in 
progress. There is, however, room for improvements of the approach described in this work, for instance, implementing CDCC or CRC 
calculations on top of our Gogny model, including 
a more accurate description of the Coulomb interaction or considering explicitly the dipole
polarizability in the deuteron scattering around the Coulomb barrier.  

        
\section*{Acknowledgement}
 
Useful discussions with J.N. De and A. Bhagwat are warmly acknowledged. 
One of the authors (X.V.) acknowledges the partial support from 
Grants No. PID2020-118758GBI00
and No. CEX2019-000918-M (through the “Unit of
Excellence María de Maeztu 2020-2023” award to ICCUB)
from the Spanish MCIN/AEI (DOI 10.13039/501100011033).

\section*{Appendix}
As it has been pointed out in the main text, the imaginary part of MOPG for light particles is too strong compared to the predictions of the KD 
model and the phenomenological models fitted to describe the elastic scattering of this kind of particles \cite{han06,pang09,su15}. This fact is, 
actually, a direct consequence of the strong absorption in the nucleon-nucleus Gogny model, as it was pointed out in \cite{lopez21}. To cure this 
unwanted effect, we renormalize with energy-dependent factors the real and imaginary parts of the central term of the nucleon-nucleus Gogny 
potential  as well as its corresponding spin-orbit contribution, in a similar way as was done in Ref.\cite{bauge98} with the 
Jeukene, Lejeune and Mahaux optical potential. 

\begin{table}[ht]
\begin{center}
\caption{Numerical values of the parameters of the energy-dependent renormalization functions $M_\alpha$, $M_\beta$ y $M_\gamma$ for 
proton-nucleus reactions described with the  MOPG}
\begin{tabular}{|c|c|c|}
\multicolumn{3}{c}{$M_\alpha(E,\lambda_k)=\lambda_1E^2+\lambda_2E+\lambda_3$} \\
\hline
$\lambda_1$ & $\lambda_2$ & $\lambda_3$ \\
\hline
4.555$\times 10^{-5}$ & -9.5351$\times 10^{-3}$ & 1.1801 \\
\hline
\multicolumn{3}{c}{$M_\beta(E,\mu_k)=\mu_1E^2+\mu_2E+\mu_3$} \\
\hline
$\mu_1$ & $\mu_2$ & $\mu_3$ \\
\hline
1.0069$\times 10^{-4}$ & -1.1196$\times 10^{-2}$ & 0.8684 \\
\hline
\multicolumn{3}{c}{$M_\gamma(E,\delta_k)=\delta_1E^2+\delta_2E+\delta_3$} \\
\hline
$\delta_1$ & $\delta_2$ & $\delta_3$ \\
\hline
-1.0329$\times 10^{-5}$ & 1.7524$\times 10^{-3}$ & 0.6147 \\
\hline
\end{tabular}
\label{table:tableP}
\end{center}
\end{table}

\begin{table}[ht]
\begin{center}
\caption{Numerical values of the parameters of the energy-dependent renormalization functions $M_\alpha$, $M_\beta$ y $M_\gamma$ for neutron-nucleus reactions described with the  MOPG} 
\begin{tabular}{|c|c|c|c|}
\multicolumn{4}{c}{$M_\alpha(E,\lambda_k)=\lambda_1e^{-\lambda_2E}+\lambda_3$} \\
\hline
$\lambda_1$ & $\lambda_2$ & $\lambda_3$ & \\
\hline
0.3019 & 0.2577 & 1.0858 & \\
\hline
\multicolumn{4}{c}{$M_\beta(E,\mu_k)=\frac{\mu_1E^2+\mu_2E+\mu_3}{E+\mu_4}$} \\
\hline
$\mu_1$ & $\mu_2$ & $\mu_3$ & $\mu_4$ \\
\hline
0.23856 & -19.3728 & 575.8818 & 383.7557 \\
\hline
\multicolumn{4}{c}{$M_\gamma(E,\delta_k)=\delta_1e^{\delta_2E}+\delta_3$} \\
\hline
$\delta_1$ & $\delta_2$ & $\delta_3$ & \\
\hline
4.4806$\times 10^{-2}$ & 0.1 & 0.5466 & \\
\hline
\end{tabular}
\label{table:tableN}
\end{center}
\end{table}

To this end we first select a set of forty-five proton-nucleus and twenty-two neutron-nucleus reactions on targets of $^{40}$Ca, $^{56}$Fe, 
$^{90}$Zr, $^{120}$Sn and $^{208}$Pb for which experimental angular distributions of elastic
scattering in the energy ranges between 10 and 100 MeV (for protons) and between 5 and 26 MeV (for neutrons) exist in the EXFOR Database 
\cite{exfor}. Next for each reaction and each energy, we determine the coefficients $\alpha$, $\beta$
and $\gamma$, which multiply the real central part, the imaginary central part and the spin-orbit contribution of the Gogny optical potential, 
in such a way that the renormalized potential minimize the relative {\it rms} build up as
\begin{equation}
rms^2=\frac{1}{N_1}\sum_{i=1}^{N}\left[\frac{\sigma_{{exp}_i}-\sigma_{th}(\alpha,\beta,\gamma)_i}{\sigma_{{exp}_i}}\right]^2,
\label{rms1}
\end{equation}
where $\sigma_{{exp}_i}$ are the $N_1$ experimental values of the differential cross-section for each considered reaction taken from \cite{exfor} 
and $\sigma_{th}$ the corresponding theoretical values computed with the renormalized potential.
We fit these two set of forty-five (protons) and twenty-two (neutrons) of \{$(\alpha,\beta,\gamma)$\} pseudo-data by suitable analytical functions,
 which allow determining the energy-dependent factors that have to be used to renormalize the Gogny optical potential for any nucleon-nucleus 
reaction. To this end, we chose the renormalization functions $M_{\alpha}(E,\lambda_k)$, $M_{\beta}(E,\mu_k)$ and $M_{\gamma}(E,\delta_k)$  given 
in Table \ref{table:tableP} for protons and in Table \ref{table:tableN} for neutrons. These functions depend on the energy of the projectile and on
 3-4 parameters, which are determined by minimizing the relative {\it rms} between each set of pseudo-data and corresponding fitting functions. 
For example, for the set of pseudo-data \{$\alpha_i$\} and the fitting function $M_{\alpha}(E,\lambda_k)$, we obtain the $\lambda_1$, $\lambda_2$ 
and $\lambda_3$ parameters by minimizing
\begin{equation}
rms^2=\frac{1}{N_2}\sum_{i=1}^N\left[\frac{\alpha_i-M_\alpha(E_i,\lambda_k)}{\alpha_i}\right]^2,
\label{rms2}
\end{equation}
where now the sum runs over the considered proton-nucleus (neutron-nucleus) reactions, $N_2$ being the total number of reactions, i.e. forty-five 
for protons and twenty-two for neutrons.


\section{References}
\begin{thebibliography}{9}
\bibitem{lopez21}
J. Lopez Mora\~na and X. Vi\~nas, J.of Phys. G48, 035104 (2021).
\bibitem{hodgson63}
P.E. Hodgson, The Optical Model of Elastic Scattering(Oxford, Clarendon) 1963.
\bibitem{becchetti69}
F.D. Becchetti and G.W. Greenlees, Phys. Rev. 182, 1190 (1969).
\bibitem{varner91}
R.L. Varner, W.J. Thomson, T.L. Abbe, E.J. Ludwig and T.B. Clegg, Phys. Rep. 201, 57 (1991).
\bibitem{koning03}
A.J. Koning and J.P. Delaroche, Nucl. Phys. A713, 231 (2003).
\bibitem{bell59}
J.S. Bell and E.J. Squires, Phys. Rev. Lett. 3, 96 (1959).
\bibitem{jeukene74}
J.P. Jeukenne, A. Lejeune and C. Mahaux, Phys. Rev. C10, 1391 (1974).
\bibitem{jeukene76}
J.P. Jeukenne, A. Lejeune and C. Mahaux, Phys. Rep. 25C, 83 (1976).
\bibitem{jeukene77a}
J.P. Jeukenne, A. Lejeune and C. Mahaux, Phys. Rev. C15, 10 (1977).
\bibitem{jeukene77b}
J.P. Jeukenne, A. Lejeune and C. Mahaux, Phys. Rev. C16, 80 (1977).
\bibitem{bauge98}
E. Bauge, J.P. Delaroche and M. Girod, Phys. Rev. C58, 1118 (1998).
\bibitem{sinha75} 
B. Sinha, Phys. Rep. 20, 1 (1975).
\bibitem{brieva77}
F.A. Brieva, J.R. Rook,
Nucl. Phys. A271, 299 (1977); Nucl. Phys. A271, 317 (1977); Nucl. Phys. A297, 206 (1978).
\bibitem{arellano90}
H.F. Arellano, F.A. Brieva and W.G. Love, Phys. Rev. C41, 2188 (1990):
Phys. Rev. C50, 02480 (1994); Phys. Rev. C52, 301 (1995).
\bibitem{arellano07}
H.F. Arellano and E. Bauge, Phys. Rev. C76, 014613 (2007); Phys. Rev. C84, 034606 (2011).
\bibitem{aguayo08}
F.J. Aguayo and H.F. Arellano,
Phys. Rev. C78, 014608 (2008).
\bibitem{amos00}
K. Amos, P.J. Dortmans, H.V. von Geram, S. Karataglidis and J. Raynal,
Ad.in Nucl. Phys. 25, 275 (2000).
\bibitem{khoa02}
Dao Tien Khoa, E. Khan, G. Col\`o and Nguyen Van Giai,
Nucl. Phys. A706, 61 (2002).
\bibitem{loan19}
Doan Thi Loan, Dao Tien Khoa and Nguyen Hoang Phuc,
J. of Phys. G47, 035106 (2019).
\bibitem{feshbach92}
H. Feshbach, Theoretical Nuclear Physics, Vol.II, (Wiley New York) 1992.
\bibitem{satchler83}
G.R. Satchler, Direct Nuclear Reactions (Clarendon Press, Oxford, 1983).
\bibitem{satchler79}
G.R. Satchler and W.G. Love,
Phys. Rep. 55, 183 (1979).
\bibitem{brandon97}
M.E. Brandan and G.R. Satchler,
Phys. Rep. 285, 143 (1977).
\bibitem{li09}
Xiaohua Li, Haixia An and Chonghai Cai,
Eur. Phys. J. A39, 255 (2009).
\bibitem{pang11}
D.Y. Pang, Y.L. Ye and F.R. Xu,
Phys. Rev. C83, 064619 (2011). 
\bibitem{vautherin72}
D. Vautherin and D.M. Brink, Phys. Rev. C5, 626 (1972).
\bibitem{decharge80}
J. Decharg\'e and D. Gogny, Phys. Rev. C21, 1568 (1980).
\bibitem{shen81}
Shen Quingbiao, Zhang Jingshang, Tien Ye, Ma Zhongyu and Zhuo Yizhong, Z. Phys. 303, 69 (1981).
\bibitem{shen09}
Qing-biao Shen, Yin-lu Han and Hai-rui Guo Phys. Rev. C80, 024604 (2009).
\bibitem{xu14}
Yong-li Xu, Hai-rui Guo, Yin-lu Han and Qing-biao Shen, J.of Phys. G41, 015101 (2014).
\bibitem{pilipenko10}
V.V. Pilipenko, V.I. Kuprikov and A.P. Soznik, Phys. Rev. C81, 044614 (2010).
\bibitem{pilipenko12}
V.V. Pilipenko and V.I. Kuprikov, 
Phys. Rev. C86, 064613 (2012).
\bibitem{pilipenko15}
V.V. Pilipenko and V.I. Kuprikov, 
Phys. Rev. C92, 014616 (2015).
\bibitem{kuprikov12}
V.I. Kuprikov, V.V. Pilipenko and A.P. Soznik,
Phys. Atom. Nuclei 75, 832 (2012).
\bibitem{guo10}
H.-R. Guo, Y.-L. Xu, Y.-L. Han and Q.-B. Shen,
Phys. Rev. C81, 044617 (2010).
\bibitem{guo14}
H.-R. Guo, Y.-L. Xu, Y.-L. Han and Q.-B. Shen,
Nucl. Phys. A922, 84 (2014).
\bibitem{guo09}
H.-R. Guo, Y. Zhang, Y.-L. Han and Q.-B. Shen,
Phys. Rev. C79, 064601 (2009).
\bibitem{guo11}
H.-R. Guo, Y.-L. Xu, H.-Y. Liang, Y.-L. Han and Q.-B. Shen,
Phys. Rev. C83, 064618 (2011).
\bibitem{egashira14}
K. Egashira, K. Minomo, M. Toyokawa, T. Matsumoto and M. Yahiro,
Phys. Rev. C89, 064611 (2014).
\bibitem{kuprikov16}
V.I. Kuprikov and V.V. Pilipenko,
Phys. Rev. C94, 064612 (2016).
\bibitem{soubbotin00}
V.B. Soubbotin and X. Vi\~nas,
 Nucl. Phys. A665, 291 (2000)
\bibitem{soubbotin03}
V.B. Soubbotin, V.I. Tselyaev and X. Vi\~nas, 
Phys. Rev. C67, 014324 (2003).
\bibitem{lopez23} 
J. Lopez Mora\~na and X. Vi\~nas,
J.of Phys. G50, 045108 (2023).
\bibitem{watanabe58}
S. Watanabe,
Nucl. Phys. 8, 484 (1958).  
\bibitem{rawitscher74}
G.H. Rawitscher,
Phys. Rev. C9, 2210 (1974).
\bibitem{keeley08}
N. Keeley and R.S. Mackintosh,
Phys. Rev. C77, 054603 (2008).
\bibitem{mackintosh19}
R.S. Mackintosh and N. Keeley,
Phys. Rev. C100, 064613 (2019).
\bibitem{keeley20}
N. Keeley and R.S. Mackintosh,
Phys. Rev. C102, 064611 (2020).
\bibitem{keeley23}
N. Keeley and R.S. Mackintosh,
Phys. Rev. C107, 034602 (2023).
\bibitem{chau06}
P. Chau Huu-Tai,
 Nucl. Phys. A771, 56 (2006).
\bibitem{andres94}
M.V. Andr\'es, J. G\'omez-Camacho and M.A. Nagarajan,
 Nucl. Phys. A579, 243 (1994).
\bibitem{moro99}
A.M. Moro and J. G\'omez-Camacho,
 Nucl. Phys. A648, 141 (1999).
\bibitem{austern87}
N. Austern, Y. Iseri, M. Kamimura, M. Kawai, G. Rawitscher and M. Yahiro,
Phys. Rep. 154, 125 (1987).
\bibitem{berger91}
J.F. Berger, M. Girod and D. Gogny, 
Comput. Phys. Commun. 63, 365 (1991).
\bibitem{cea}
CEA web page www-phynu.cea.fr
\bibitem{pillet17}
N. Pillet and S. Hilaire, 
Eur. Phys. J. A53, 10 (2017).
\bibitem{chappert08}
F. Chappert, M. Girod and S. Hilaire, 
Phys. Lett. B668, 420 (2008).
\bibitem{goriely09}
S. Goriely, S. Hilaire, M. Girod and S. Peru, 
Phys. Rev. Lett. 102, 242501 (2009).
\bibitem{lei19}
J. Lei and A. Moro, 
Phys. Rev. Lett. 122, 042503 (2019).
\bibitem{baur76}
G. Baur,
Phys. Lett. B178, 135 (1986).
\bibitem{bhagwat09}
A. Bhagwat and Y.K. Gambhir,
J.of Phys. G36, 025105 (2014
\bibitem{roubos06}
D. Roubos, A. Pakou, N. Alamanos and K. Rusek,
Phys. Rev. C73, 051603(R) (2006).
\bibitem{han06}
Yinlu Han, Yuyang Shi and Qingbiao Shen,
Phys. Rev. C74, 044615 (2006).
\bibitem{pang09}
D.Y. Pang, P. Roussel-Chomaz, H. Savajols, R.L. Varner and R. Wolski,
Phys. Rev. C79, 024615 (2009). 
\bibitem{su15}
Xin-Wu Su and Yin-Lu Han
Int. J. Mod. Phys. E24, 01550092 (2015). 
\bibitem{exfor}
Experimental Nuclear Reaction Data (EXFOR) Database Version of 2023-02-13.
\bibitem{hatanaka80} 
K. Hatanaka, K. Imai, S. Kobayashi, T. Matsusue, M. Nakamura, K. Nisimura, T. Noro, H. Sakamoto, 
H. Shimizu and J. Shirai,
Nucl. Phys. A340, 93 (1980).
\bibitem{kiss76}
A. Kiss, O. Aspelund, G. Hrehuss, K.T. Kn\"ofle, M. Rogge, U. Schwinn, Z. Seres, P.Turek and
C. Mayer-Boricke,
Nucl. Phys. A262, 1 (1976).
\bibitem{baumer01}
C. B\"aumer, R. Bassini, A.M. van der Berg, D. De Frenne, D. Frekers, M. Hagemann, V.M. Hannen, M.N. Harakeh,
J. Heyse, M.A. de Huu, E. Jacobs, M. Mielke, S. Rakers, R. Schmidt, H. Sohlbach and H.J. W\"ortche,
Phys. Rev. C63, 037601 (2001).
\bibitem{england87}
J.B.A. England et al.
Nucl. Phys. A475, 422 (1987).
\bibitem{hyakutake78}
M. Hyakutake at al.,
Nucl. Phys. A311, 161 (1978).
\bibitem{hauser69}
G. Hauser, R. L\"ohken, H. Rebel, G. Schatz, G.W. Schweimer and J. Speth,
Nucl. Phys. A128, 81 (1969).
\bibitem{rebel72a}
H. Rebel, G.W. Schweimer, G. Schatz, J. Speth,  R. L\"ohken, G. Hauser, D. Habs and H. Klewe-Nebenius,
Nucl. Phys. A182, 145 (1972).
\bibitem{rebel72b}
K. Rebel,  R. L\"ohken, G.W. Schweimer,  G. Schatz and G. Hauser,
Z. Phys 256, 258 (1972).
\bibitem{gils75}
H.J. Gils, H. Rebel, G. Nowicki, A. Ciocamel, D. Hartmann, H. Klewe-Nebenius and H. Wisshak,
J.of Phys. G1, 344 (1975).
\bibitem{england82}
J.B.A. England et al.
Nucl. Phys. A388, 173 (1982).
\bibitem{auce96}
A. Auce, R.F. Carlson, A.J. Cox, A. Ingemarsson, R. Johansson, P.U. Renberg, O. Sundberg and G. Tibell,
Phys. Rev. C53, 2919 (1996).
\bibitem{ingemarsson01}
A. Ingemarsson, G.J. Arendse, A. Auce, R.F. Carlson, A.A. Cowley, A.J. Cox, S.V. F\"ortsch, R. Johansson,
B.R. Karlson, M. Lantz, J. Peavy, J.A. Stander, G.F. Steyn and G. Tibell,
Nucl. Phys. A696, 3 (2001).
\bibitem{ingemarsson00}
A. Ingemarsson, J. Nyberg,  P.U. Renberg, O. Sundberg, R.F. Carlson, A.J. Cox, A. Auce, R. Johansson, G. Tibell,
Dao T. Khoa and R.E. Warner
Nucl. Phys. A676, 3 (2000).
\end{thebibliography}
\end{document}